\documentclass[english,aps,prl, twocolumn,superscriptaddress]{revtex4-1}
\usepackage[utf8]{inputenc}
\usepackage{amsmath}
\usepackage{amssymb}
\usepackage{graphicx}
\usepackage{hyperref}
\usepackage{siunitx}
\usepackage{placeins}
\usepackage{braket}
\usepackage{babel}
\usepackage{mathtools}
\usepackage{xspace}

\usepackage[utf8]{inputenc}
\usepackage[T1]{fontenc}
\usepackage{lmodern}
\usepackage{graphics}
\usepackage{xcolor}
\usepackage{dsfont}
\usepackage{rotating}
\usepackage{float}
\usepackage{commath}

\begin{document}
\title{Photon-assisted resonant Andreev reflections: Yu-Shiba-Rusinov and Majorana states}

\author{Sergio Acero Gonz\'alez}
\affiliation{\mbox{Dahlem Center for Complex Quantum Systems and Fachbereich Physik, Freie Universit\"at Berlin, 14195 Berlin, Germany}}

\author{Larissa Melischek}
\affiliation{\mbox{Dahlem Center for Complex Quantum Systems and Fachbereich Physik, Freie Universit\"at Berlin, 14195 Berlin, Germany}}

\author{Olof Peters}
\affiliation{\mbox{Fachbereich Physik, Freie Universit\"at Berlin, 14195 Berlin, Germany}}

\author{Karsten Flensberg}
\affiliation{\mbox{Center for Quantum Devices, Niels Bohr Institute, University of Copenhagen, DK-2100 Copenhagen, Denmark}}
\affiliation{\mbox{Dahlem Center for Complex Quantum Systems and Fachbereich Physik, Freie Universit\"at Berlin, 14195 Berlin, Germany}}

\author{Katharina J. Franke}
\affiliation{\mbox{Fachbereich Physik, Freie Universit\"at Berlin, 14195 Berlin, Germany}}

\author{Felix von Oppen}
\affiliation{\mbox{Dahlem Center for Complex Quantum Systems and Fachbereich Physik, Freie Universit\"at Berlin, 14195 Berlin, Germany}}

\begin{abstract}
Photon-assisted tunneling frequently provides detailed information on the underlying charge-transfer process. In particular, the Tien-Gordon approach and its extensions predict that the sideband spacing in bias voltage is a direct fingerprint of the number of electrons transferred in a single tunneling event. Here, we analyze photon-assisted tunneling into subgap states in superconductors in the limit of small temperatures and bias voltages where tunneling is dominated by resonant Andreev processes and does not conform to the predictions of simple Tien-Gordon theory. Our analysis is based on a systematic Keldysh calculation of the subgap conductance and provides a detailed analytical understanding of photon-assisted tunneling into subgap states, in excellent agreement with a recent experiment. We focus on tunneling from superconducting electrodes and into Yu-Shiba-Rusinov states associated with magnetic impurities or adatoms, but we also explicitly extend our results to include normal-metal electrodes or other types of subgap states in superconductors. In particular, we argue that photon-assisted Andreev reflections provide a high-accuracy method to measure small, but nonzero energies of subgap states which can be important for distinguishing conventional subgap states from Majorana bound states. 

\end{abstract}

\maketitle

\section{Introduction}

At subgap temperatures and voltages, charge transfer between conventional superconductors typically occurs by multi-electron processes. Transfer of Cooper pairs is responsible for Josephson currents flowing between superconductors \cite{Josephson1962} and leaves the superconductors in their ground state. Cooper pairs can also be extracted from, injected into, or transferred between superconductors with the simultaneous generation of quasiparticles \cite{Schrieffer1963,Andreev1964}. In these processes -- termed multiple Andreev reflections -- electrons impinging on one of the superconducting electrodes are reflected as holes, while a Cooper pair is transmitted into the superconductor. As a result, one or several Cooper pairs are transferred between the superconductors while generating a pair of quasiparticles \cite{Schrieffer1963,Octavio83,Averin1995,Bratus1995}. 

\begin{figure}[t]
\centering
\includegraphics[width=0.9\columnwidth]{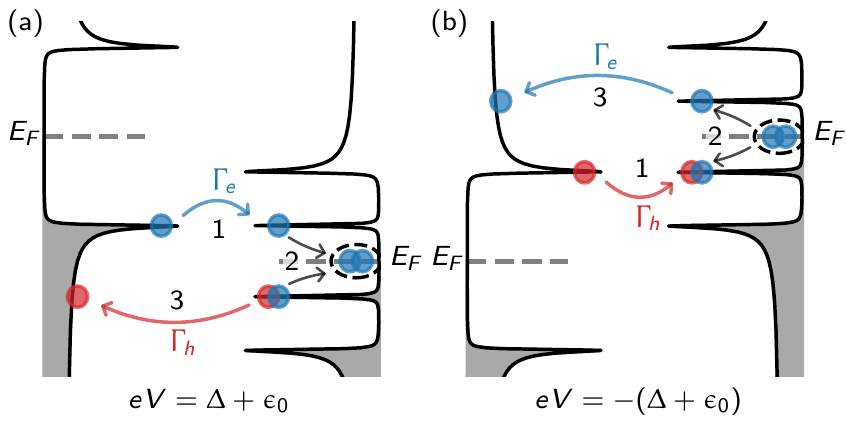}
\caption{Resonant Andreev reflections via YSR bound states in superconductor-superconductor junctions at threshold (schematic, no high-frequency radiation). (a) For positive bias voltages, an electron (blue) tunnels from the coherence peak of the source (tip) into the positive-energy YSR state, which then forms a Cooper pair with another electron while creating a hole (red) in the negative-energy YSR state. Finally, the hole tunnels back into the source. (b) For negative bias voltages, a hole tunnels from the coherence peak of the drain (tip) into the negative-energy YSR state. A Cooper pair breaking up in the source (substrate) will then compensate the hole and occupy the positive-energy YSR state, followed by electron tunneling into the drain (tip). The processes at positive and negative bias both create a pair of quasiparticles in the tip (left superconductor) and generate or break up a Cooper pair in the substrate (right superconductor). 
\label{Fig1_Ac}}
\end{figure}

At subgap voltages, single-electron transmission is possible only due to thermally excited quasiparticles. In tunnel junctions, these processes can compete with two-electron tunneling since the latter are of higher order in the tunneling amplitude and hence exponentially suppressed. The interplay of single-electron and two-electron tunneling can be elucidated in scanning-tunneling-spectroscopy experiments where the junction resistance is readily changed by orders of magnitude, thereby tuning the relative importance of these two tunneling processes. In a recent experiment \cite{Ruby2015b}, this was done for a system in which tunneling was resonantly enhanced by in-gap Yu-Shiba-Rusinov (YSR) states associated with a magnetic adatom. Single-electron tunneling dominated for large tip-substrate distances, where tunneling processes are slow compared to inelastic processes coupling the YSR state to the quasiparticle continuum. In contrast, the tunnel current was predominantly carried by two-electron processes at smaller tip-substrate distances where the tunneling processes are fast. These resonant two-electron processes -- which we term resonant Andreev reflections, see Fig.\ \ref{Fig1_Ac} -- transfer a Cooper pair into the substrate while generating a pair of quasiparticles in the tip. The nature of these processes was further elucidated by a subsequent experiment \cite{Peters2020} which aimed at distinguishing single-electron and two-electron tunneling through YSR states by means of photon-assisted tunneling in the presence of high-frequency (HF) radiation \cite{foot10}. Here, we develop a comprehensive theory of the tunneling processes as well as the resulting intriguing and nontrivial patterns of photon-assisted sidebands. 

Photon-assisted tunneling constitutes a powerful method to probe the nature of charge transfer. The absorption and emission of photons leads to the appearance of sidebands in the conductance both in the absence \cite{Dayem1962,Roychowdhury2015} and in the presence \cite{Kouwenhoven1994,Blick1995,Nakamura1999,Meyer2007,Gramich2015,Zanten2019} of Coulomb blockade. 
Frequently, the spacing of the sidebands in bias voltage as well as their modulation as a function of the amplitude of the HF radiation directly reveal the amount of charge that is transferred in an elementary tunneling event \cite{Platero2004}. The theory of such processes goes back to the classic work of Tien and Gordon \cite{Tien1963}, and their early results on single-electron transfer between superconductors has been extended in multiple directions. In many situations, one finds Tien-Gordon-like relations 
\begin{equation}
   G(V) = \sum_n J_n^2\left( \frac{keV_{\mathrm{HF}}}{\hbar\Omega}\right) G^{(0)}(V + n\hbar\Omega/ke),
\label{TienGordonLike}
\end{equation}
which express the junction conductance $G(V)={\mathrm d}I/{\mathrm d}V$ in the presence of HF irradiation in terms of the junction conductance $G^{(0)}(V)$ without HF radiation. Here, $k$ denotes the number of electrons transferred in an elementary tunneling event, $\Omega$ is the frequency of the HF radiation, and $V_{\mathrm{HF}}$ its amplitude. The conductance is a sum over sidebands spaced in bias voltage by $\hbar\Omega/ke$, whose strength is controlled by the Bessel functions $J_n$. The oscillations of the Bessel functions as a function of their argument imply a characteristic modulation of the sideband intensity as a function of $V_{\mathrm{HF}}$.

Such relations have been shown to describe not only photon-assisted sidebands of the coherence peaks \cite{Tien1963}, but also incoherent Josephson tunneling near zero bias \cite{Falci1991} or multiple Andreev reflections \cite{Zimmermann1996}. In the context of scanning-tunneling-microscopy (STM) experiments, these Tien-Gordon expressions were found to describe the sidebands of the coherence peaks \cite{Peters2020,Kot2020}, the Josephson peak \cite{Roychowdhury2015,Peters2020,Kot2020}, as well as multiple Andreev peaks \cite{Peters2020,Kot2020}. To understand these relations, it is convenient to measure energies in both source and drain from the respective chemical potentials. In this representation, the Hamiltonian $H_T$ describing tunneling from source to drain involves a time-dependent phase factor (see Sec.\ \ref{sec:model} for details)
\begin{equation}
    e^{-i\phi(\tau)} = e^{-i\{\frac{eV}{\hbar}\tau+\frac{eV_{\rm HF}}{\hbar\Omega}\sin \Omega\tau\}},
\end{equation}
which accounts for the change in energy of the tunneling electrons due to the voltage bias across the junction. The amplitude for transferring multiple electrons can be obtained from higher-order terms in the Born series for the $T$-matrix, $T= H_T + H_T G_0 H_T + \ldots$. While in general, the unperturbed Green function $G_0$ is nonlocal in time, it is effectively local on the scale of $\Omega^{-1}$ when the energy of the virtual intermediate states is large compared to the energy transfer from the HF radiation. In this case, the factors of $e^{-i\phi(\tau)}$ from the various tunneling terms simply combine into a single factor $e^{-ik\phi(\tau)}$, and a simple Fermi golden rule calculation leads to Eq.\ (\ref{TienGordonLike}). This argument applies to incoherent Cooper pair tunneling as well as multiple Andreev reflections for plain superconducting electrodes as long as $\hbar\Omega, eV_{\mathrm{HF}} \ll \Delta$. 

It is clear that this reasoning does not extend to resonant Andreev reflections via YSR states where the amplitude for tunneling is sharply peaked in energy due to the bound state. Indeed, we find that photon-assisted resonant Andreev reflections exhibit rich physics that is qualitatively different from the Tien-Gordon-like expression (\ref{TienGordonLike}). Moreover, while in many cases, the tunneling between superconductors can be described in low-order perturbation theory in the tunneling Hamiltonian, this is generally not the case for resonant Andreev reflections \cite{Ruby2015b}. This is because the broadening of the bound-state resonance can be dominated by the tip-substrate tunneling, thus requiring one to treat tunneling to all orders in perturbation theory. We show that nevertheless, one can develop an analytical theory for photon-assisted resonant Andreev reflections. Our theory is in excellent agreement with a recent experiment \cite{Peters2020} on tunnel junctions formed between a superconducting substrate with a magnetic adatom and a superconducting STM tip. 

Resonant Andreev reflections are an important tunneling process not only for YSR states, but also for other subgap states in superconductors. In particular, they dominate tunneling into Majorana bound states, where they are predicted to lead to a universal zero-bias conductance of $2e^2/h$ for tunneling from a normal-metal lead \cite{Law2009,Flensberg2010}. This has been at the focus of a large number of experiments \cite{Lutchyn2018} and a recent measurement shows evidence for this quantized conductance \cite{Nichele2017}. Our theory for photon-assisted resonant Andreev reflections is readily adapted to include tunneling into Majorana bound states \cite{Tang2015,Zanten2019}, and we find that photon-assisted tunneling can be an important tool to differentiate Majorana bound states from other subgap states. This is particularly true for tunneling from superconducting tips which were repeatedly used for improved resolution in Majorana experiments on chains of magnetic adatoms \cite{NadjPerge2014,Ruby2015,Feldman2017,Ruby2017}. Since tunneling into a Majorana bound state leaves behind an unpaired electron in the superconducting tip, it leads to two symmetric Majorana peaks at bias voltages $eV=\pm \Delta$, where $\Delta$ denotes the superconducting gap of the tip \cite{Peng2015}. This should be contrasted with tunneling into a conventional subgap state with a small, but nonzero energy $\epsilon_0$, which appears as differential-conductance peaks at $eV=\pm(\Delta+\epsilon_0)$. Thus, the small energy of the subgap state can only be extracted from experiment as a difference of two much larger energies, the position of the resonance peak in ${\mathrm d}I/{\mathrm d}V$ and the superconducting gap of the tip. This is inherently prone to errors and requires an accurate determination of the tip gap. We find that in photon-assisted tunneling, the subgap energy appears directly as a spacing between resonant peaks in the spectrum, even for a superconducting tip. Moreover, these splittings appear in differential-conductance maps with high multiplicity, which effectively enhances the ability to resolve closely-spaced peaks. 

Building on a brief review of subgap tunneling processes between pristine superconductors in Sec.\ \ref{sec:pristine}, we begin in Sec.\ \ref{sec:RAP} with a summary of our central results and the basic physical picture for photon-assisted resonant Andreev reflections via YSR states in superconductor-superconductor junctions. The model and some basic formalism for tunneling between superconductors are set up in Sec.\ \ref{sec:model}. Our central analytical results for photon-assisted resonant Andreev reflections are then derived in Sec.\ \ref{sec:RARsec}. We first discuss a diagonal approximation in Sec.\ \ref{sec:diag} which is in excellent agreement with experiment and exact theoretical results, and apply this approach to photon-assisted resonant Andreev reflections in junctions of normal metals and superconductors with YSR state (Sec.\ \ref{sec:normalmetal}) as well as superconductor-superconductor junctions (Sec.\ \ref{sec:suptip}), giving a firm theoretical basis to the physical discussion in Sec.\ \ref{sec:phys}. We then derive and discuss the exact solution in Sec.\ \ref{sec:exact}. While the bulk of the paper is concerned with YSR states, many results carry over rather directly to Majorana bound states, as discussed in Sec.\ \ref{sec:Majorana}. We conclude in Sec.\ \ref{sec:conclusions}.

\section{Physical discussion}
\label{sec:phys}

Before embarking on the detailed technical derivation of the photon-assisted tunneling current, we begin with a physical discussion. We include a review of standard results for tunnel junctions between superconductors to provide a backdrop for resonant Andreev reflections in junctions with YSR states. A corresponding discussion of resonant Andreev reflections via Majorana bound states can be found in Sec.\ \ref{sec:Majorana}.

\subsection{Review of photon-assisted tunneling processes between pristine superconductors}
\label{sec:pristine}

Single-electron tunneling between superconductors leaves behind an unpaired electron in the source and injects an unpaired electron into the drain. Each of these electrons requires a minimal excitation energy equal to the superconducting gap  $\Delta$ (assumed equal for source and drain superconductors for simplicity). Thus, single-electron tunneling becomes possible at voltages $e|V|>2\Delta$. The BCS singularity of the superconducting density of states leads to coherence peaks in the differential conductance at the threshold voltages $eV=\pm 2\Delta$. In the presence of an $ac$ field with frequency $\Omega$, the tunneling electrons not only gain energy $eV$ due to the bias voltage, but also emit or absorb photons \footnote{It has recently been argued \cite{Kot2020} that these processes should not be described in terms of photons. Here, we nevertheless follow this long-standing and established terminology as it does not really lead to confusion.}. Then, the threshold condition for single-electron tunneling becomes $eV + n\hbar\Omega = \pm 2\Delta$, where the integer $n$ is positive for photon absorption and negative for photon emission, and one obtains a set of coherence peaks displaced in voltage by multiples of the photon energy $\hbar\Omega/e$ \cite{Tien1963, Dayem1962}. The number of emitted or absorbed photons per tunneling event is bounded by the maximal energy $eV_{\mathrm {HF}}$ that the tunneling electrons can exchange with the $ac$ field, where $V_{\mathrm {HF}}$ denotes the amplitude of the $ac$ bias across the junction. This implies that coherence-peak sidebands are limited to $|n|\lesssim n_{\mathrm{max}}=eV_{\mathrm {HF}}/\hbar\Omega$ and thus observable in the voltage range $2\Delta - eV_{\mathrm {HF}} \lesssim e|V| \lesssim 2\Delta + eV_{\mathrm {HF}}$. 

Current can also flow at subgap voltages due to multiple Andreev reflections. Electrons with subgap energies impinging on the source or drain superconductor are reflected as holes, with a Cooper pair transferred into the superconductor (or vice versa). Then, the required excitation energy of $2\Delta$ for the two generated quasiparticles can be acquired in the course of multiple traversals across the junction, and the threshold condition becomes $meV=2\Delta$, where the (positive) integer $m$ denotes the number of junction traversals and thus the number of electrons transmitted into the drain superconductor. In the presence of the $ac$ field, photons can be emitted or absorbed in the tunneling process, and the threshold condition becomes $meV+n\hbar\Omega=2\Delta$. The spacing of the photon sidebands in voltage is then given by $\hbar\Omega/me$ and directly reflects the number of transferred electrons per tunneling process. Specifically, the lowest multiple Andreev process with $m=2$ has a threshold voltage of $eV=\Delta$ without $ac$ field, transmits a Cooper pair into the drain, and has sidebands with a voltage spacing of $\hbar\Omega/2e$ \cite{Uzawa2005, Peters2020, Kot2020}.

In the vicinity of zero bias, current flow between superconductors occurs via Cooper pair tunneling. This leads to a zero-bias peak in the differential conductance, reflecting that Cooper pair tunneling does not excite either of the superconducting electrodes 
\footnote{The finite width of the Josephson peak as a function of bias voltage is associated with dissipation into the modes of the electromagnetic environment of the tunnel junction.}. The $ac$ field splits this Josephson peak into sidebands. The Cooper pairs gain an energy $2eV$ due to the applied bias and $n\hbar\Omega$ due to the photon field. Thus, these sidebands occur at $eV=n\hbar\Omega/2$, exhibiting half the spacing in bias voltage compared to single-electron processes and the same spacing as the $m=2$ Andreev processes \cite{Cuevas2002}. The tunneling Cooper pairs change their energy at most by $2eV_{\mathrm {HF}}$ due to the $ac$ field. Consequently, the Josephson peaks are limited to $|n|\lesssim n_{\mathrm{max}}=2eV_{\mathrm {HF}}/\hbar\Omega$ and visible in the voltage range $- eV_{\mathrm {HF}} \lesssim eV \lesssim eV_{\mathrm {HF}}$ \cite{Nakamura1999, Naaman2001, Roychowdhury2015, Peters2020, Kot2020}. 

At nonzero temperatures, there are additional single-electron processes even at subgap voltages which originate from thermally excited quasiparticles. The latter lead to a peak in the differential conductance when the coherence peaks of the two superconductors align. This causes a zero-bias peak when source and drain have gaps of the same magnitude, and more generally a peak at $eV=\pm |\Delta_1-\Delta_2|$, when the superconductors have different gaps \cite{Ternes2006, Franke2011}.

\subsection{Resonant Andreev processes via YSR states}
\label{sec:RAP}

Magnetic adatoms induce bound states -- known as YSR states \cite{Yu1965,Shiba1968,RusinovA.I.1969,Balatsky2006}  -- within the superconducting gap which can be individually probed by scanning tunneling spectroscopy \cite{Yazdani1997,Ji2008,Franke2011, Menard2015, Balatsky2006, Heinrich2018}. The YSR states induce additional resonances in the tunneling conductance at subgap voltages $e|V|<2\Delta$. At zero temperature, the subgap current cannot be carried by single electrons. Due to the absence of bulk states at these energies, single electrons cannot leave the junction region. Instead, the dominant current-carrying process is an Andreev process closely related to the lowest multiple Andreev process discussed above with $m=2$ \cite{Deacon2010, Ruby2015b, Randeria2016, Lee2017, Brand2018, Farinacci2018}. This process -- termed resonant Andreev reflection -- is best viewed as a (coherent) multistep process. First consider the situation when electrons are tunneling from the tip into the substrate (positive bias voltage, see Fig.\ \ref{Fig1_Ac}(a) for a schematic representation). In this case, the tunneling amplitude involves the following steps. An electron from the tip initially tunnels into the positive-energy YSR state. Subsequently, the electron combines with an electron in the substrate to form a Cooper pair, allowing the charge to exit the junction region and leaving behind a hole in the negative-energy YSR state. Finally, this hole tunnels back into the tip. 

This process must satisfy two conditions to be energetically allowed, one each for electron and hole \cite{Ruby2015b,Peters2020}. The electron tunneling process virtually occupies the YSR state of energy $\epsilon_0$ and leaves an unpaired electron behind in the tip, and is thus allowed when $eV >\Delta +\epsilon_0$. The hole tunneling process injects a hole into the quasiparticle continuum and thus requires $eV >\Delta -\epsilon_0$. Since $\epsilon_0>0$, the condition for hole tunneling is automatically satisfied whenever the condition for electron tunneling is met. Thus, resonant Andreev reflection induces a peak in the differential conductance at the threshold bias voltage $eV=\Delta+\epsilon_0$ of electron tunneling.  In contrast, there is no peak at the hole threshold $eV=\Delta-\epsilon_0$ since the electron process is not yet energetically allowed.  

\begin{figure}[t]
\centering
\includegraphics[width=0.95\columnwidth]{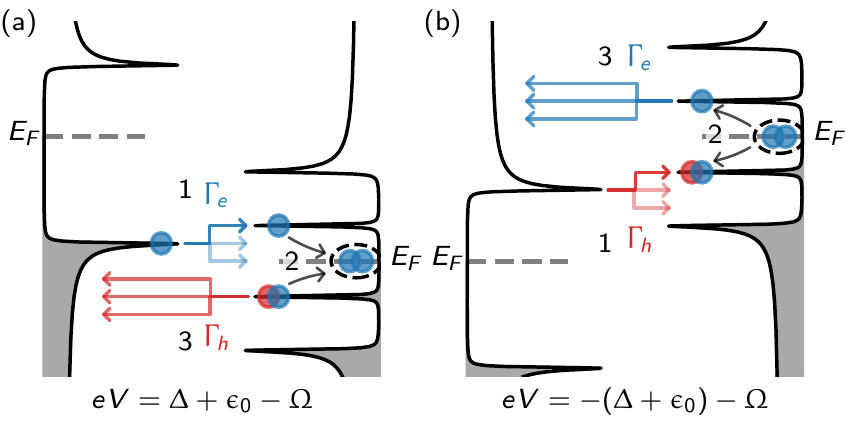}
\caption{Photon-assisted resonant Andreev reflections via YSR bound states in superconductor-superconductor junctions (schematic). Panel (a) shows the process for positive bias voltages, panel (b) for negative biases. The basic process is as in Fig.\ \ref{Fig1_Ac}. The high-frequency radiation (frequency $\Omega$) allows electrons and holes to change their energy by multiples of $\hbar\Omega$. The number of relevant sidebands is limited by the maximal energy $eV_{\rm HF}$ that the tunneling electrons and holes can gain or lose due to the high-frequency field and therefore grows linearly in $V_{\rm HF}$.  
\label{Fig3_Ac}}
\end{figure}

\begin{figure*}[t]
\centering
\includegraphics[width=1.0\textwidth]{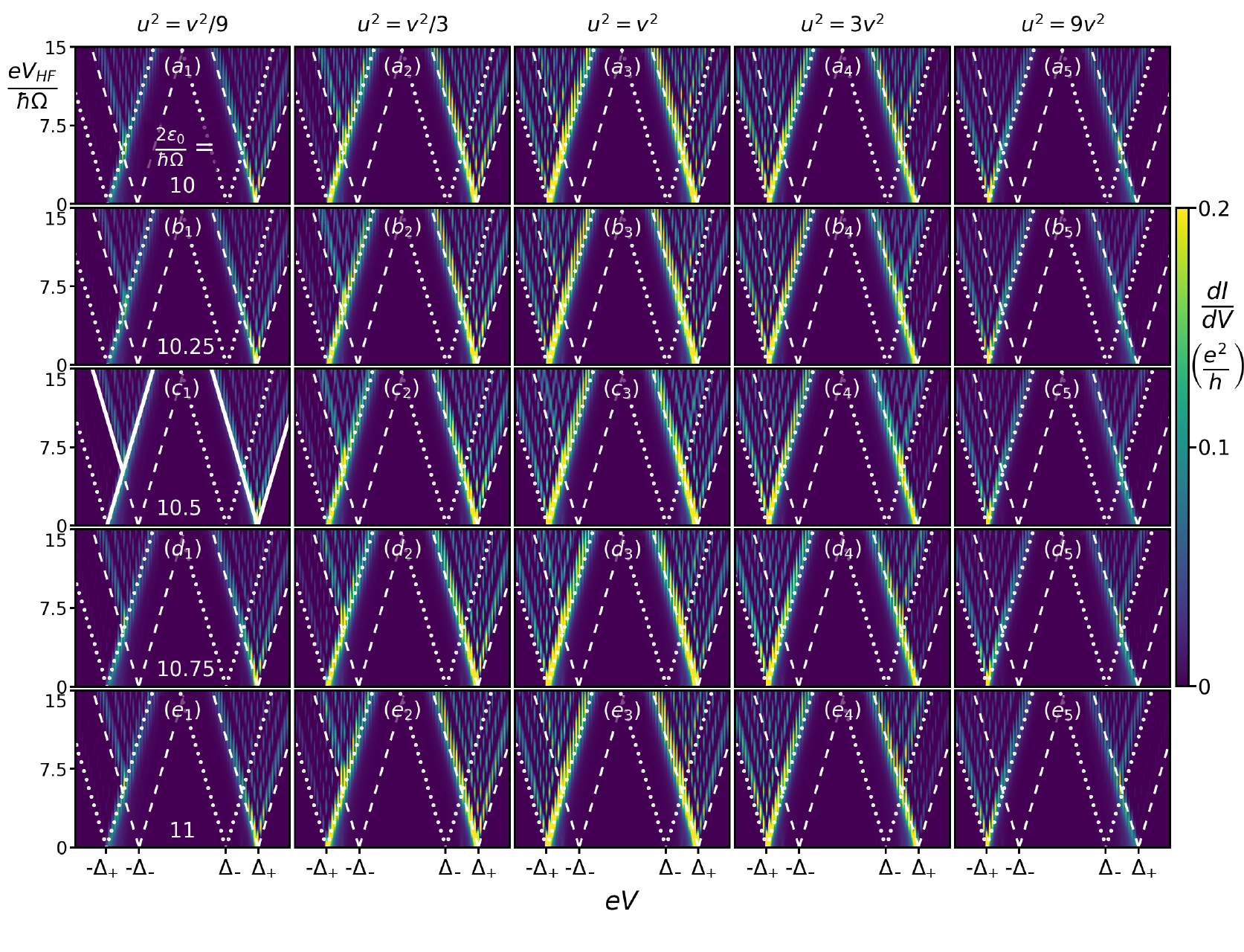}
\caption{Differential conductance (color scale) as a function of bias voltage $eV$ and amplitude $eV_{\rm HF}$ of the high-frequency radiation for tunneling from a superconducting tip into a YSR state via resonant Andreev reflections. The panels differ in the ratio between electron and hole wavefunctions $u$ and $v$ (left to right) and in the ratio between YSR state energy $\epsilon_0$ and photon energy $\hbar\Omega$ (top to bottom). Numerical values are indicated in the figure. These results are obtained for $u^2+v^2$ fixed to the same value for all panels. The regions with electron and hole sidebands (see Eqs.\ (\ref{elside}) and (\ref{ellsideneg}) as well as (\ref{holside}) and (\ref{hosideneg}), respectively) are indicated by white dashed and dotted V-shapes, respectively, centered at $e|V|=\Delta_\pm= \Delta \pm \epsilon_0$. Notice the appearance of V and Y-shaped regions, as highlighted in panel (c1). For a detailed discussion, see Sec.\ \ref{sec:RAP}. Parameters: $\Omega/\Delta = 0.05$, $\nu_0|t| = 0.04$.
\label{Fig2_Ac}}
\end{figure*}

In the presence of an $ac$ field, both the electron and the hole can emit or absorb photons during tunneling, cf.\ Fig.\ \ref{Fig3_Ac}, and the energetic conditions become $eV >\Delta +\epsilon_0 + n\hbar\Omega$ for the electron and $eV >\Delta -\epsilon_0 + m\hbar \Omega$ for the hole. Correspondingly, there are two sets of sidebands in the differential conductance, one at $eV =\Delta +\epsilon_0 + n\hbar\Omega$ due to the condition for electron tunneling and another at $eV =\Delta -\epsilon_0 + m\hbar \Omega$ due to the condition for hole tunneling.
Electron and hole can both gain or lose a maximal energy of $eV_{\mathrm{ HF}}$ due to the $ac$ field. Thus, the electron sidebands are restricted to the voltage region  
\begin{eqnarray}   
       \Delta + \epsilon_0 - eV_{\mathrm{ HF}} \lesssim  eV \lesssim \Delta + \epsilon_0 + eV_{\mathrm{ HF}} ,
       \label{elside}
\end{eqnarray}
and the hole sidebands to 
\begin{eqnarray}   
       \Delta - \epsilon_0 - eV_{\mathrm{ HF}} \lesssim  eV \lesssim \Delta - \epsilon_0 + eV_{\mathrm{ HF}} .
       \label{holside}
\end{eqnarray}
These V-shaped regions are indicated in Fig.\ \ref{Fig2_Ac} as dashed (electrons) and dotted (holes) lines. The sidebands are observable only as long as both electron and hole tunneling are allowed. For positive bias, this limits them to the voltage range (\ref{elside}) for electron sidebands. Within this region, only electron sidebands are observed for $eV > \Delta - \epsilon_0 + eV_{\mathrm{ HF}}$, i.e., outside the dotted V shape for hole sidebands. Both electron and hole sidebands contribute for $eV < \Delta - \epsilon_0 + eV_{\mathrm{ HF}}$, which corresponds to the region lying within both dashed and dotted V shapes.

At negative bias voltages, there is a corresponding process in which a hole tunnels from the tip into the negative-energy YSR state, a Cooper pair breaks up and occupies both YSR states at positive and negative energies, and finally, an electron tunnels back from the positive-energy YSR state into the tip, see Fig.\ \ref{Fig1_Ac}(b). In this process, the hole sidebands are limited to the region
\begin{eqnarray}   
       -(\Delta + \epsilon_0 + eV_{\mathrm{ HF}}) \lesssim  eV \lesssim -(\Delta + \epsilon_0 - eV_{\mathrm{ HF}}), 
       \label{hosideneg}
\end{eqnarray}
while electron sidebands can appear in the region 
\begin{eqnarray}   
       -(\Delta - \epsilon_0 + eV_{\mathrm{ HF}}) \lesssim  eV \lesssim -(\Delta - \epsilon_0 - eV_{\mathrm{ HF}}) .
       \label{ellsideneg}
\end{eqnarray}
In the absence of high-frequency radiation, it is now the electron process that is above threshold whenever the hole process is, and sidebands can only be observed within the hole region given by Eq.\ (\ref{hosideneg}). 

Based on our full theoretical results (see Sec.\ \ref{sec:RARsec}), Fig.\ \ref{Fig2_Ac} exhibits the differential conductance as a function of both bias voltage $V$ and amplitude $V_{\mathrm{HF}}$ of the $ac$ field. From top to bottom, the panels differ in the ratio between YSR energy $\epsilon_0$ and photon energy $\hbar\Omega$. From left to right, the panels differ in the ratio between electron and hole wavefunctions $u$ and $v$, respectively, evaluated at the tip position. First consider the column of central panels for equal amplitudes of electron and hole wavefunctions, $|u|^2=|v|^2$. The differential conductance exhibits pronounced V-shapes centered at $eV=\pm(\Delta+\epsilon_0)$. At positive bias, this V-shape reflects the region with electron sidebands given in Eq.\ (\ref{elside}), at negative biases the region with hole sidebands given in Eq.\ (\ref{hosideneg}).

These panels also show clear evidence for the importance of both the electron and the hole condition. The panels in Fig.\ \ref{Fig2_Ac} delineate the V-shaped regions both for electron tunneling (dashed white lines) and for hole tunneling (dotted white lines). The sideband structure within the outer V-shaped regions differs markedly between the overlap region of the two V-shapes and the region outside the inner V-shape. Generically, one observes a larger number of sidebands within the overlap region where both electron and hole thresholds contribute. Only when $2\epsilon_0$ is commensurate with $\hbar\Omega$, electron and hole thresholds coincide and the sidebands in the overlap region appear brighter, but not more numerous. 

Strikingly, the inner arms of the V-shapes appear more pronounced than the outer ones. This can be understood as follows. The sidebands appear brighter in the differential conductance, if the YSR resonance is sharp. The width of the YSR resonance is controlled by the electron and hole tunneling rates. Along the inner arm, one of the tunneling processes is just barely setting in, so that the width is considerably smaller than along the outer arm, where both electron and hole tunneling are fully allowed. 

The patterns depend strongly on the ratio between electron and hole wavefunctions. Consider now the leftmost column of panels in Fig.\ \ref{Fig2_Ac}, for which the hole wavefunction is considerably larger than the electron wavefunction, $|u|^2=|v|^2/9$. While one still observes a V-shaped region of sidebands for positive bias voltages, the region takes on a Y-shape for negative biases. Since the hole wavefunction is much larger, hole tunneling rates are intrinsically larger than electron tunneling rates. In this case, electron tunneling is effectively the rate-limiting process (see Sec.\ \ref{sec:suptip} for a careful discussion of this statement) and electron thresholds are considerably more pronounced than hole thresholds. Thus, sidebands are only observed within the overlap region. The only exception is the ``stem'' of the Y-shape along which hole tunneling just sets in and is still comparable in magnitude to electron tunneling. The situation is analogous in the rightmost column in Fig.\ \ref{Fig2_Ac}, for which the electron tunneling rate is typically much larger than the hole tunneling rate and a (reflected) Y-like shape now appears at positive bias voltages. 

Since one set of sidebands dominates for strongly asymmetric electron and hole wavefunctions, the sidebands no longer depend sensitively on the commensurability between $2\epsilon_0$ and $\hbar\Omega$, but appear with a regular voltage spacing of $\hbar\Omega$. In view of the simple Tien-Gordon relation in Eq.\ (\ref{TienGordonLike}), this seemingly suggests that the underlying tunneling process is a single-electron process. Nevertheless, resonant Andreev reflections transfer electron pairs into the substrate superconductor and should be viewed as a single coherent process. This emphasizes that photon-assisted resonant Andreev reflections do not conform to the predictions of a simple Tien-Gordon approach. 

\begin{figure}[t]
\centering
\includegraphics[width=0.95\columnwidth]{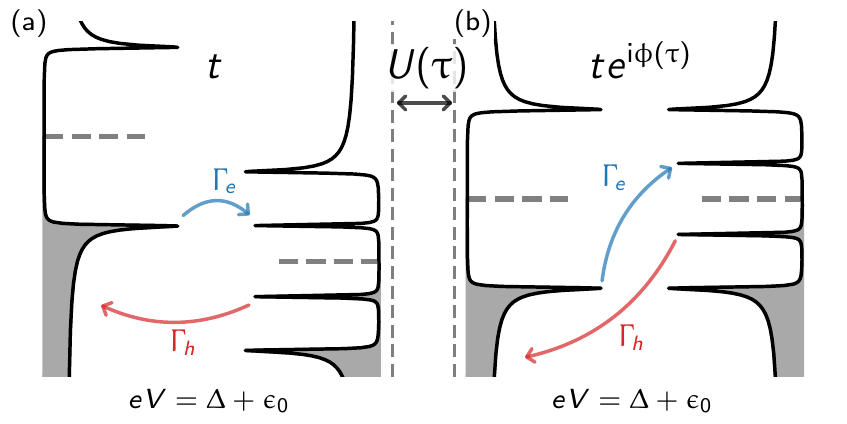}
\caption{Alternative representations of electron and hole tunneling in the presence of a bias voltage: (a) Left and right superconductors have chemical potentials which are shifted relative to one another by the applied bias voltage $eV$. In this representation, tunneling is horizontal, leaving the energy unchanged. (b) Alternatively, a time-dependent unitary transformation, see Eq.\ (\ref{timedependentU}), shifts the chemical potentials of left and right superconductor such that they become equal, and tunneling of electrons and holes is associated with an energy transfer equal to $eV$. We use the representation in panel (a) for figures, but the calculations (and their description) are systematically performed using the representation in panel (b). 
\label{Fig4_Ac}}
\end{figure}

\section{Model}
\label{sec:model}

We consider a junction involving a superconducting tip and substrate (or other kinds of superconducting electrodes) with Hamiltonians $\tilde H_L$ and $\tilde H_R$, respectively. Electrons can tunnel between tip and substrate as described by the tunneling Hamiltonian
\begin{equation}
\tilde H_T=\sum_\sigma\left[t{c}^\dagger_{L,\sigma}({\bf R}) {c}_{R,\sigma}({\bf R})+{\rm h.c.}\right],
\end{equation}
where ${c}^\dagger_{\alpha,\sigma}({\bf r})$ creates an electron at position ${\bf r}$ and spin $\sigma$ in the tip ($\alpha=L$) or the substrate ($\alpha=R$) and ${\bf R}$ denotes the position of the tip. The Hamiltonian 
\begin{equation}
   \tilde H = \tilde H_L + \tilde H_R + \tilde H_T
\end{equation}
measures energy on an absolute scale and conserves the total particle number $N=N_L+N_R$. The time-dependent bias $V(\tau)$ between tip and substrate is included by holding tip and substrate at different chemical potentials $\mu_L$ and $\mu_R$, 
\begin{equation}
   eV(\tau) = \mu_L-\mu_R,
\end{equation} 
and is the sum of an applied $dc$ voltage $V$ and an $ac$ voltage 
\begin{equation}
   V_{\rm ac}(\tau)=V_{\rm HF}\cos(\Omega\tau)
\end{equation}
generated by the radiation field of frequency $\Omega$ \cite{Tien1963}.

To apply the usual BCS mean-field description of the superconducting tip and substrate, we perform a time-dependent canonical transformation (setting $\hbar = 1$)
\begin{equation}
   U(\tau) = \exp\{ i \int_0^{\tau} {\mathrm d}\tau^\prime [\mu_L(\tau^\prime) N_L + \mu_R(\tau^\prime) N_R]  \},
\label{timedependentU}
\end{equation}
so that single-particle energies in tip and substrate are measured from the respective chemical potentials $\mu_L$ and $\mu_R$. The transformed Hamiltonian $H=U\tilde H U^\dagger-iU\partial_\tau U^\dagger$ takes the form
\begin{equation}
   H = (\tilde H_L -\mu_L N_L) + (\tilde H_R - \mu_R N_R) + U\tilde H_T U^\dagger.   
\end{equation}
Here, we used that $\tilde H_L$ and $\tilde H_R$ conserve $N_L$ and $N_R$, so that $U \tilde H_\alpha U^\dagger = \tilde H_\alpha$. Then, the time dependence enters only through the transformed tunneling Hamiltonian $H_T = U\tilde H_T U^\dagger$ with
\begin{equation}
H_T=\sum_\sigma\left[te^{i\phi(\tau)}{c}^\dagger_{L,\sigma}({\bf R}){c}_{R,\sigma}({\bf R})+{\rm h.c.}\right],
\end{equation}
where the tunneling amplitude $t$ acquires a time-dependent phase
\begin{equation}
   {\phi(\tau)}=eV\tau+\frac{eV_{\rm HF}}{\Omega}\sin(\Omega\tau)
\end{equation} 
as a result of the canonical transformation. While the time-independent $\tilde H_T$ conserves the energy of the tunneling electron or hole, the time-dependent $H_T$ changes the energy due to both, the applied $dc$ and $ac$ biases. This corresponds to different representations of the same tunneling process as illustrated in Fig.\ \ref{Fig4_Ac}.

In the transformed Hamiltonian, we can now make the usual BCS mean field approximation for both $H_L = \tilde H_L -\mu_L N_L$ and $H_R = \tilde H_R -\mu_R N_R$. The unperturbed Hamiltonian $H_0=H_L + H_R$ can then be written as 
\begin{eqnarray}
H_0&=&\sum_{{\bf k},\alpha} \sum_{\sigma} \left[\xi_{{\bf k},\alpha} {c}^\dagger_{\alpha,\mathbf{k}\sigma}{c}_{\alpha,\mathbf{k}\sigma}+\left(\Delta{c}^\dagger_{\alpha,\mathbf{k}\uparrow}{c}^\dagger_{\alpha,-\mathbf{k}\downarrow}+{\rm h.c.}\right) \right] \nonumber\\
&& \,\,\,\,\,\,\,\,+\sum_{{\bf k},{\bf k}^\prime}\sum_\sigma (K-JS\sigma){c}^\dagger_{R,{\bf k}\sigma}{c}_{R,{\bf k}^\prime\sigma},
\end{eqnarray}
where $\xi_{{\bf k},\alpha}=\epsilon_{\bf k}-\mu_\alpha$ denotes the normal-state dispersion for lead $\alpha=L,R$ and $c^\dagger_{\alpha,\mathbf{k}\sigma}$ creates an electron with momentum ${\mathbf k}$. The superconducting gap $\Delta$ is taken to be identical for tip and substrate. A magnetic adatom with spin ${\bf S}$ is located at the origin and modeled as a classical impurity which couples to the substrate electrons via potential scattering of strength $K$ and exchange coupling $J$. The spin quantization axis of the electrons is chosen parallel to the impurity spin. 

The current operator $I=-e\dot{N}_L$ takes the form
\begin{eqnarray}
   I&=&  -ie [H_T,N_L]  \nonumber \\
    &=& ie\sum_\sigma\left(te^{i\phi(\tau)}{c}^\dagger_{L,\sigma}({\bf R}){c}_{R,\sigma}({\bf R}) -{\mathrm 
    {h.c.}} \right)
\end{eqnarray}
and the current becomes 
\begin{equation}
I(\tau)=e\mathrm{Tr}\left\{\tau_z \left[\hat{t}(\tau)G_{RL}^<(\tau,\tau)-G^<_{LR}(\tau,\tau)\hat{t}^*(\tau) \right]\right\}.
\label{current1}
\end{equation}
Here, we have expressed the expectation values in terms of the lesser Green function in Nambu space,
\begin{equation}
  G^<_{\alpha\beta}(\tau_1,\tau_2) = i \left(  \begin{array}{cc}  
  \langle c^\dagger_{\beta\uparrow}(\tau_2) c_{\alpha\uparrow}(\tau_1)\rangle & 
  \langle c_{\beta\downarrow}(\tau_2) c_{\alpha\uparrow}(\tau_1)\rangle \\ 
  \langle c^\dagger_{\beta\uparrow}(\tau_2) c^\dagger_{\alpha\downarrow}(\tau_1)\rangle & 
  \langle c_{\beta\downarrow}(\tau_2) c^\dagger_{\alpha\downarrow}(\tau_1)\rangle 
  \end{array}  \right),
\end{equation}
introduced the hopping matrix
\begin{equation}
 \hat{t}(\tau)= \left( \begin{matrix} te^{i\phi(\tau)} & 0 \\
0 & -t^*e^{-i\phi(\tau)}
\end{matrix} \right),
\end{equation}
and used the Pauli matrix $\tau_z$ in Nambu space. Here and in the following, electron operators (as well as Green functions and self energies) without momentum or position labels refer to the tip position ${\bf R}$. 

Writing Dyson equations for the Keldysh Green function and using the Langreth rules, the lesser Green functions can be written as 
\begin{equation}
\begin{split}
G^<_{LR}&=(g_L\hat{t}G_R)^<=g^<_L\hat{t}G^a_R+g^r_L\hat{t}G_R^<\\
G^<_{RL}&=(G_R \hat{t}^*g_L)^<=G_R^< \hat{t}^*g_L^a+G_R^r \hat{t}^*g^<_L.
\end{split}
\label{Langreth}
\end{equation} 
The superscripts $r$ and $a$ denote retarded and advanced Green functions. The bare Green function (in Nambu space) of tip or substrate in the absence of tunneling is denoted as $g_\alpha$ ($\alpha=L,R$), while
the Green function of the substrate which accounts for the tip-substrate tunneling through a self energy 
\begin{equation}
  \Sigma_R(\tau,\tau^\prime) = \hat{t}^*(\tau) g_L(\tau,\tau^\prime)\hat{t}(\tau^\prime)
  \label{selfenergy}
\end{equation}
takes the form $G_R=[g_R^{-1}-\Sigma_R]^{-1}$. 

Inserting Eqs.\ (\ref{Langreth}) into the expression (\ref{current1}) for the current, we find
\begin{eqnarray}
   I(\tau) &=& e\int {\mathrm d}\tau^\prime\, {\rm Tr} \left\{ \tau_z \left[ 
                G^<_R(\tau,\tau^\prime) \Sigma_R^a(\tau^\prime,\tau) 
                \right. \right. \nonumber\\
                && + G^r_R(\tau,\tau^\prime) \Sigma_R^<(\tau^\prime,\tau)
                - \Sigma^<_R(\tau,\tau^\prime) G_R^a(\tau^\prime,\tau) \nonumber\\
                && \left.\left.- \Sigma^r_R(\tau,\tau^\prime) G_R^<(\tau^\prime,\tau)\right]\right\}  .
                \label{basiccurrent}
\end{eqnarray}
Here, we used that the hopping matrix $\hat t$ commutes with $\tau_z$.

\section{Resonant Andreev reflections}
\label{sec:RARsec}

While the YSR states resonantly enhance Andreev processes in the substrate, no such enhancement occurs for Andreev reflections in the tip. For this reason, we effectively neglect the latter. In this approximation, there are no multiple Andreev reflections, and the dominant processes contributing to the subgap conductance involve a single resonant Andreev reflection in the substrate. We can implement this approximation by neglecting the off-diagonal contributions to the Nambu Green function $g_L$ of the tip when computing the self energy $\Sigma_R$. In this approximation, $g_L$ is proportional to the unit matrix (see App.\ \ref{app:gL} for details).

To compute the self energy $\Sigma_R$ within this approximation, we note that 
\begin{equation}
   e^{i\phi(\tau)} = \sum_{n=-\infty}^\infty J_n\!\left(\frac{eV_{\mathrm {HF}}}{\Omega}\right) e^{i(eV +  
   n\Omega)\tau},
\label{Besselexp}
\end{equation}
where $J_n(x)$ denotes a Bessel function. Inserting this into Eq.\ (\ref{selfenergy}), we obtain
\begin{eqnarray}
   &&\Sigma_R(\tau,\tau^\prime) = |t|^2 \sum_{n,m} J_n\!\left(\frac{eV_{\mathrm {HF}}}{\Omega}\right)
   J_m\!\left(\frac{eV_{\mathrm {HF}}}{\Omega}\right) 
   \nonumber\\
   &&\,\,\,\,\times
    e^{-i(eV+n\Omega)\tau\tau_z}  g_L(\tau-\tau^\prime) e^{i(eV+m\Omega)\tau^\prime\tau_z}.
\end{eqnarray}
This expression can be viewed as a sum of a diagonal ($n=m$) and an off-diagonal ($n\neq m$) contribution,
\begin {equation}
   \Sigma_R = \Sigma_R^0 + \Sigma_R^1
\end{equation}
with 
\begin{eqnarray}
 &&\Sigma^0_R(\tau,\tau^\prime) = |t|^2 \sum_{n} J_n^2({eV_{\mathrm {HF}}}/{\Omega}) 
 \nonumber \\
 &&\,\,\,\,\,\,\times e^{-i(eV+n\Omega) \tau \tau_z}  g_L(\tau-\tau^\prime) e^{i(eV+n\Omega)\tau^\prime\tau_z}.
 \label{Sigdiag}
\end{eqnarray}
The calculation simplifies significantly when retaining only the diagonal self energy $\Sigma_R^0$. We find that this is frequently an excellent approximation. For this reason, we first discuss this simplified situation (referred to below as diagonal approximation) before presenting the more general case. 

\subsection{Diagonal approximation}
\label{sec:diag}

\subsubsection{Derivation}
\label{sec:calc}

Within the diagonal approximation, the self energy is only a function of the difference $\tau-\tau^\prime$ of its time arguments and thus diagonal in frequency representation. Then, the exponential factors in Eq.\ (\ref{Sigdiag}) effectively act as translation operators and we obtain 
\begin{eqnarray}
 \Sigma^0_R(\omega) = |t|^2 \sum_{n} J_n^2({eV_{\mathrm {HF}}}/{\Omega}) 
  g_L(\omega\!-\!(eV \!+\! n\Omega)\tau_z). \,\,\,\,\hbox{}
\end{eqnarray}
Here, the frequency argument of the Green function $g_L$ reflects that due to bias voltage and $ac$ field, electrons (holes) propagating in the substrate lose (gain) an energy $eV+n\Omega$  when tunneling into the tip. 

As we are considering subgap energies in the substrate, we only retain the contribution to the substrate Green function which originates from the YSR state with energy $\epsilon_0$. Then, the retarded and advanced Green functions become (see App.\ \ref{app:SubG} for details)
\begin{equation}
    G^{r/a}_R(\omega) =   \psi \frac{1}{\omega-\epsilon_0-\Lambda(\omega) \pm 
    \frac{i}{2}\Gamma(\omega)} \psi^\dagger .
    \label{GR}
\end{equation} 
Here,  $\psi^T=(u ,v)$ denotes the Bogoliubov-deGennes wavefunction of the positive-energy YSR state at the tip position ${\bf R}$ and we separated the retarded and advanced self energy projected onto the YSR state
\begin{equation}
   \tilde \Sigma^{0,r/a}_R(\omega)=\psi^\dagger \Sigma^0_R(\omega)\psi = \Lambda(\omega) \mp 
   \frac{i}{2}\Gamma(\omega)
\end{equation}
into real and imaginary parts.

The projected self energy takes the explicit form
\begin{eqnarray}
  &&\tilde \Sigma^0_R(\omega) = |t|^2 \sum_{n} J_n^2({eV_{\mathrm {HF}}}/{\Omega}) 
 \nonumber \\
 &&\times \left\{ |u|^2 g_L(\omega\!-\!(eV \!+\! n\Omega)) + |v|^2 g_L(\omega\!+\!(eV \!+\! n\Omega)) \right\}. \,\,\,\,\,\,\,\,\,
 \label{tildeSigmaExp}
\end{eqnarray}
Using Eq.\ (\ref{gLra}) in App.\ \ref{app:gL}, the imaginary part $\Gamma(\omega)$ is given by
\begin{eqnarray}
    &&\Gamma(\omega) = \sum_{n} J_n^2({eV_{\mathrm 
    {HF}}}/{\Omega}) \nonumber\\ 
    &&\,\,\,\,\,\,\,\,\,\,\,\times [\Gamma_e(\omega\!-\!(eV \!+\! n\Omega))+ 
    \Gamma_h(\omega\!+\!(eV \!+\! n\Omega))],
\label{GammaSideBands}
\end{eqnarray}
which combines contributions to the broadening of the YSR state due to photon-assisted tunneling of electrons and holes into the tip. Here, we defined the electron and hole tunneling rates
\begin{eqnarray}
  \Gamma_e(\omega) &=& 2\pi |u|^2 |t|^2  \nu(\omega) = \gamma_e \nu(\omega)/\nu_0,  \\
  \Gamma_h(\omega) &=& 2\pi |v|^2 |t|^2  \nu(\omega) = \gamma_h \nu(\omega)/\nu_0
\end{eqnarray}
with the BCS density of states
\begin{equation}
   \nu(\omega) = \nu_0  \frac{|\omega|}{\sqrt{\omega^2-\Delta^2}}
  \, \theta(|\omega| - \Delta)
\label{BCSDOS}
\end{equation}
of the tip. Here, $\nu_0$ is the normal-state density of states per spin direction and we introduced the tunneling rates $\gamma_e=2\pi |u|^2|t|^2\nu_0$ and $\gamma_h=2\pi |v|^2|t|^2\nu_0$ for a normal-state tip. Similarly, the real part of the self energy becomes
\begin{eqnarray}
  &&\Lambda(\omega) = -\pi\nu_0 |t|^2 \sum_{n} J_n^2({eV_{\mathrm {HF}}}/{\Omega}) 
 \nonumber \\
 &&\times \left\{ \frac{|u|^2 [\omega\!-\!(eV \!+\! n\Omega)]}{\sqrt{\Delta^2 -[\omega\!-\!(eV \!+\! n\Omega)]^2}}\theta(\Delta-|\omega\!-\!(eV \!+\! n\Omega)|) \right.\,\,\,\,\,\,\,\,\,\,\, \nonumber \\
 && \left. + \frac{|v|^2 [\omega\!+\!(eV \!+\! n\Omega)]}{\sqrt{\Delta^2 -[\omega\!+\!(eV \!+\! n\Omega)]^2}}\theta(\Delta-|\omega\!+\!(eV \!+\! n\Omega)|) \right\},
\end{eqnarray}
describing a (frequency-dependent) renormalization of the energy of the YSR state. 

The lesser self energy can be expressed by inserting Eq.\ (\ref{gLless}) into Eq.\ (\ref{tildeSigmaExp}). This yields
\begin{eqnarray}
    &&\tilde\Sigma_R^<(\omega) = i \sum_{n} J_n^2({eV_{\mathrm 
    {HF}}}/{\Omega}) \nonumber\\ 
    &&\,\,\,\,\,\,\,\,\,\,\,\times [\Gamma_e(\omega\!-\!(eV \!+\! n\Omega)) n_F(\omega\!-\!(eV \!+\! n\Omega)) \nonumber\\
   && \,\,\,\,\,\,\,\,\,\,\,+ \Gamma_h(\omega\!+\!(eV \!+\! n\Omega))] n_F(\omega\!+\!(eV \!+\! n\Omega))].
\end{eqnarray}
This also yields the lesser Green function 
\begin{equation}
   G_R^<(\omega) = \psi \frac{\tilde\Sigma_R^<(\omega)}{[\omega-\epsilon_0-\Lambda(\omega)]^2 + 
    \frac{1}{4}\Gamma^2(\omega)}\psi^\dagger
\end{equation}
of the substrate using the relation (\ref{Glessrlessa}) in App.\ \ref{app:SubG}.

Within the diagonal approximation, we can then express Eq.\ (\ref{basiccurrent}) for the current in frequency representation, 
\begin{eqnarray}
   I(\tau) &=& e\int \frac{{\mathrm d}\omega}{2\pi} {\rm Tr} \left\{ \tau_z \left[ 
                G^<_R(\omega) \Sigma_R^a(\omega)+ G^r_R(\omega) \Sigma_R^<(\omega) 
                \right. \right. \nonumber\\
                && \left.\left. - \Sigma^<_R(\omega) G_R^a(\omega)- \Sigma^r_R(\omega) G_R^<(\omega)\right]\right\}  .
\end{eqnarray}
This can be written in the alternative form
\begin{eqnarray}
   I &=& e \int \frac{{\mathrm d}\omega}{2\pi}{\rm Tr} \left\{ \tau_z \left[ G_R^r(\omega) (\Sigma_R^r(\omega)-\Sigma_R^a(\omega)) G_R^a(\omega)  
   \Sigma_R^<(\omega)
  \right. \right. \nonumber\\
                && \,\,\,\,\,\,\,\,\left.\left. - G_R^r(\omega) \Sigma_R^<(\omega) G_R^a(\omega) 
                \left( \Sigma_R^r(\omega) - \Sigma^a_R(\omega) \right) \right]\right\} 
                \label{currentbasic}
\end{eqnarray}
using that the self energy $\Sigma_R$ is also diagonal in Nambu space and commutes with $\tau_z$ as well as the identities $G_R^r-G_R^a = G_R^r (\Sigma_R^r-\Sigma_r^a) G_R^a$ and $G_R^< = G_R^r \Sigma_R^< G_R^a$ (see Appendix \ref{app:SubG}).

With this, we are now in a position to evaluate the current in Eq.\ (\ref{currentbasic}) and obtain 
\begin{eqnarray}
  && I = 2e\int \frac{\mathrm{d}\omega}{2\pi} \sum_{n,m}J_n^2({eV_{\mathrm {HF}}}/
  {\Omega})J_m^2({eV_{\mathrm {HF}}}/{\Omega}) \nonumber\\
  && \,\,\,\,\,\,\,\,\,\times \frac{\Gamma_e(\omega\!-\!(eV \!+\! n\Omega))\Gamma_h(\omega\!+\!(eV \!+\! m\Omega))}{[\omega-\epsilon_0-\Lambda(\omega)]^2 + 
    \frac{1}{4}\Gamma^2(\omega)}  \nonumber \\
   &&\,\,\,\,\,\,\,\,\,\times [n_F(\omega\!-\!(eV \!+\! n\Omega))-n_F(\omega\!+\!(eV \!+\! m\Omega))]
   \label{diagcurrentresult}
\end{eqnarray}
after some straightforward algebra. This expression generalizes the results of Ref.\ \cite{Ruby2015b} to include photon-assisted processes
and is a main result of this paper. While the current does not obey the simple Tien-Gordon relations (\ref{TienGordonLike}), the electron and hole tunneling rates by themselves behave in a Tien-Gordon-like manner. Equation (\ref{diagcurrentresult}) is not only in excellent agreement with the more complete treatment shown below, but also with recent experimental results \cite{Peters2020}. We note that we have approximated the substrate Green function by retaining the contribution of the subgap state only. As a result, Eq.\ (\ref{diagcurrentresult}) describes only those sidebands which fall within the superconducting gap. In effect, this imposes upper cutoffs on the frequency and amplitude of the HF radiation. Except for these cutoffs, the results are independent of the substrate gap.

\begin{figure}[t]
\centering
\includegraphics[width=0.95\columnwidth]{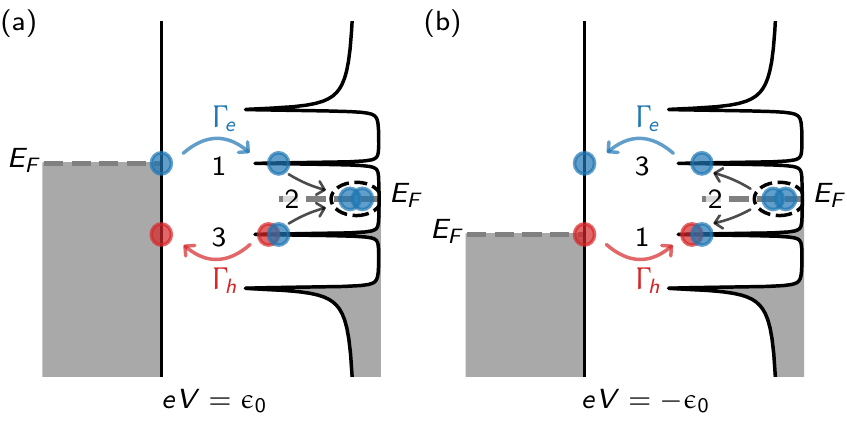}
\caption{Resonant Andreev reflections from a YSR state with a normal-state electrode for (a) positive and (b) negative polarity of the bias voltage $V$ (schematic, no high-frequency radiation).  
\label{Fig5_Ac}}
\end{figure}

\subsubsection{Normal-metal tip}
\label{sec:normalmetal}

As a first application of Eq.\ (\ref{diagcurrentresult}), consider a normal-metal tip (temporarily setting $\Delta=0$ in  the self energy) in the absence of the $ac$ field. The basic resonant Andreev reflection process in this case is illustrated in Fig.\ \ref{Fig5_Ac}. For a normal-metal tip, the self energy is purely imaginary and frequency independent, so that the bias voltage enters only into the Fermi functions. This allows one to readily evaluate the zero-temperature differential conductance,  
\begin{equation}
   \frac{\mathrm {d}I}{\mathrm {d}V} = \frac{2e^2}{h}\sum_\pm \frac{\gamma_e\gamma_h}{(eV \pm 
   \epsilon_0)^2 +(\gamma_e+\gamma_h)^2/4},
\end{equation} 
where we have reinstated Planck's constant. This yields two symmetric resonances at $eV=\pm\epsilon_0$ with peak height
\begin{equation}
    \left. \frac{\mathrm {d}I}{\mathrm {d}V}\right|_{\mathrm{peak}} = \frac{2e^2}{h} \frac{4|u|^2|v|^2}{(|u|^2+|v|^2)^2}.
\end{equation}
Thus, the peak height depends on the relative magnitudes of the electron and hole wavefunctions of the YSR state and has a maximal value of $2e^2/h$, as long as the positive- and negative-energy peaks are well separated. Specifically, the peak height becomes maximal when the electron and hole wavefunctions at the tip position are equal, $|u|^2=|v|^2$. For a YSR state with zero energy, the two peaks coalesce and the maximal peak height equals $4e^2/h$. The latter result should be compared to analogous results for Majorana bound states which give a peak conductance of $2e^2/h$ \cite{Law2009,Flensberg2010}, reflecting the fact that unlike YSR states, Majorana bound states effectively correspond to only half a conventional fermionic excitation, see also Sec.\ \ref{sec:Majorana}. 

In the presence of the $ac$ field, the zero-temperature differential conductance becomes 
\begin{eqnarray}
   \frac{\mathrm {d}I}{\mathrm {d}V} = \frac{2e^2}{h}\sum_n 
   \sum_\pm \frac{J_n^2({eV_{\mathrm {HF}}}/
  {\hbar\Omega})\gamma_e\gamma_h}{(eV +n\hbar\Omega \pm 
   \epsilon_0)^2 +\frac{(\gamma_e+\gamma_h)^2}{4}},\,\,\,\,\,\,
\label{SupTipDiag}
\end{eqnarray} 
using the Bessel-function identity $\sum_n J_n^2(x)=1$. Thus, the conductance peaks at $eV=\pm\epsilon_0$  develop sidebands whose spacings are given by the photon frequency $\hbar\Omega$ and whose amplitudes are controlled by Bessel functions. 

At finite temperatures, the peaks become convolutions of the Lorentzian with derivatives of the Fermi function in the usual manner. While the peaks are Lorentzian with a width controlled by the tunneling rates $\gamma_e$ and $\gamma_h$ at low temperatures, they cross over to derivatives of the Fermi functions at high temperatures. Here, we assume that the temperature is still sufficiently small compared to the substrate gap so that we can neglect inelastic processes which couple the YSR state to the quasiparticle continuum of the substrate. Once the latter become relevant, there is an additional contribution to the current originating from single-electron tunneling. (An experimental fingerprint of the latter is that it generically leads to asymmetric conductance peaks at $eV=\pm \epsilon_0$ \cite{Balatsky2006,Ruby2015b}.)

\subsubsection{Superconducting tip}
\label{sec:suptip}

We can now make contact with the physical discussion for a superconducting tip in Sec.\ \ref{sec:phys}. The advantages of superconducting tips are twofold. First, they enhance energy resolution owing to the sharp peak in the BCS density of states at the gap edge.  Second, when the tip is superconducting, the YSR peaks appear at $eV=\pm(\Delta+\epsilon_0)$ and the Fermi-function factor in Eq.\ (\ref{diagcurrentresult}) equals $\pm1$ to exponential precision in $T/\Delta$. Thus, the current is insensitive to temperature as long as $T\ll \Delta$, a much weaker condition than for normal-state tips where temperature should be compared to the intrinsic width of the YSR resonance \cite{Zitko2016}. 

The expression (\ref{diagcurrentresult}) clearly exhibits the coherent nature of the underlying tunneling process. Analogous to conventional resonant tunneling through a bound state \cite{Brandes1997,LevyYeyati1997}, the electron and hole tunneling rates $\Gamma_e$ and $\Gamma_h$ enter not only in the numerator, but also determine the broadening of the YSR resonance denominator. The current is then nonperturbative in the tip-substrate tunneling, and consequently {\em sublinear} in the normal-state conductance of the tunnel junction \cite{Ruby2015b}. 

An explicit evaluation of the differential conductance must take into account that as a consequence of the BCS density of states of the tip, the tunneling rates are themselves functions of $\omega$ \cite{Ruby2015b}. First consider the case without high-frequency radiation for 
positive bias voltages near the threshold $eV=\Delta+\epsilon_0$. Due to the BCS density of states, the hole contribution to the tunneling rate $\Gamma(\omega)$ in Eq.\ (\ref{GammaSideBands}) becomes of order  
\begin{equation}
  \Gamma_{h,{\textrm{thres}}} \simeq \gamma_h \sqrt{\frac{\Delta}{4\epsilon_0}}.
\label{Gammahthres}
\end{equation}
In contrast, the electron contribution becomes singular, 
\begin{equation}
   \Gamma_e(\omega-eV) \simeq \gamma_e \sqrt{\frac{\Delta}{2(\epsilon_0-\omega)}}\theta(\epsilon_0-\omega),
\label{Gammaethres}
\end{equation}
cp.\ Fig.\ \ref{Fig1_Ac}(a). 
Thus, the characteristic electron scattering rate $\Gamma_{e,{\textrm{thres}}}$ depends on whether the broadening $\Gamma$ is dominated by electron or hole tunneling. If electron tunneling dominates the broadening, we find the threshold electron tunneling rate $\Gamma_{e,{\textrm{thres}}}$ by comparing $\Gamma_e(\omega-eV)$ to the $\omega-\epsilon_0$ term in the resonance denominator of Eq.\ (\ref{diagcurrentresult}). This yields $\Gamma_{e,{\textrm{thres}}} \sim (\gamma_e^2\Delta)^{1/3}$, and Eq.\ (\ref{diagcurrentresult}) gives the peak differential conductance  
\begin{equation}
   \left. \frac{\mathrm{d} I}{\mathrm{d} V}\right|_{\mathrm{peak},+} \sim \frac{2e^2}{h}\frac{\Gamma_{h,{\textrm{thres}}}}{\Gamma_{e,{\textrm{thres}}}} \sim  \frac{ 2e^2}{h}\frac{\gamma_h \Delta^{1/6}}{\gamma_e^{2/3}\epsilon_0^{1/2}}.
\label{eggh}
\end{equation}
If on the other hand, hole tunneling dominates the broadening, $\Gamma_{h,{\textrm{thres}}}\gg\Gamma_{e,{\textrm{thres}}}$, the characteristic electron tunneling rate becomes $\Gamma_{e,{\textrm{thres}}}\sim \gamma_e(\Delta/\Gamma_{h,{\textrm{thres}}})^{1/2}$, and we find 
\begin{equation}
   \left. \frac{\mathrm{d} I}{\mathrm{d} V}\right|_{\mathrm{peak},+} \sim \frac{2e^2}{h}\frac{\Gamma_{e,{\textrm{thres}}}}{\Gamma_{h,{\textrm{thres}}}} \sim \frac{2 e^2}{h}\frac{\gamma_e \epsilon_0^{3/4}}{\gamma_h^{3/2}\Delta^{1/4}}.
\label{hgge}
\end{equation}
Details of this calculation can be found in App.\ \ref{app:dIdV}

Analogous considerations apply to negative bias voltages near the threshold $eV=-(\Delta+\epsilon_0)$, where the hole tunneling rate becomes singular at threshold while the electron tunneling rate $\Gamma_{e,{\textrm{thres}}} = \gamma_e (\Delta/4\epsilon_0)^{1/2}$ remains regular, cf.\ Fig.\ \ref{Fig1_Ac}(b). When electron tunneling dominates, $\Gamma_{e,{\textrm{thres}}} \gg \Gamma_{h,{\textrm{thres}}}$, we find the characteristic hole tunneling rate $\Gamma_{h,{\textrm{thres}}} \sim \gamma_h (\Delta/\Gamma_e)^{1/2}$ and the peak differential conductance becomes
\begin{equation}
   \left. \frac{\mathrm{d} I}{\mathrm{d} V}\right|_{\mathrm{peak},-} \sim \frac{2e^2}{h}\frac{\Gamma_{h,{\textrm{thres}}}}{\Gamma_{e,{\textrm{thres}}}} \sim  \frac{2 e^2}{h}\frac{\gamma_h \epsilon_0^{3/4}}{\gamma_e^{3/2}\Delta^{1/4}}.
\end{equation}
If on the other hand, broadening is dominated by hole tunneling, $\Gamma_{h,{\textrm{thres}}}\gg\Gamma_{e,{\textrm{thres}}}$, we find $\Gamma_{h,{\textrm{thres}}} \sim (\gamma_h^2\Delta)^{1/3}$ and 
\begin{equation}
   \left. \frac{\mathrm{d} I}{\mathrm{d} V}\right|_{\mathrm{peak},-} \sim \frac{2e^2}{h}\frac{\Gamma_{e,{\textrm{thres}}} }{\Gamma_{h,{\textrm{thres}}}} \sim  \frac{ 2e^2}{h}\frac{\gamma_e \Delta^{1/6}}{\gamma_h^{2/3}\epsilon_0^{1/2}}.
\end{equation}

In the presence of the high-frequency radiation, Eq.\ (\ref{diagcurrentresult}) exhibits photon-assisted sidebands in the differential conductance as reflected in the frequency arguments which are shifted by multiples of $\hbar\Omega$. The strength of these sidebands oscillates as a function of $V_{\mathrm{HF}}$ due to the oscillatory nature of the Bessel functions. Moreover, the Bessel functions rapidly diminish as their argument becomes larger than the index, so that the sums over $n$ and $m$ -- and  thus the photon-assisted sidebands -- are effectively restricted to the range $|n|,|m|\lesssim {eV_{\mathrm {HF}}}/{\hbar \Omega}$. It is these limits that are indicated in Fig.\ \ref{Fig2_Ac} by white (dashed and dotted) lines and reflect the fact that the tunneling electrons and holes can gain or lose at most $eV_{\rm HF}$ in energy due to the HF field. 

Equation (\ref{diagcurrentresult}) also makes the separate thresholds for electron and hole tunneling explicit, which were underlying much of our discussion in Sec.\ \ref{sec:phys}. The tunneling rates $\Gamma_{e/h}$ are proportional to the BCS density of states with its onset of density of states at $\pm\Delta$. Using the resonance denominator in Eq.\ (\ref{diagcurrentresult}) to replace $\omega$ by the bound-state energy $\epsilon_0$ in the electron and hole tunneling rates $\Gamma_e(\omega\!-\!(eV \!+\! n\Omega))$ and $\Gamma_h(\omega\!+\!(eV \!+\! m\Omega))$, we read off thresholds at $eV = \pm \Delta +\epsilon_0 -n\Omega$ for electron tunneling and at $eV = \pm \Delta -\epsilon_0 -m\Omega$ for hole tunneling, in agreement with the results quoted in Sec.\ \ref{sec:RAP} (up to the irrelevant sign of the integers $n,m$). 

We now use Eq.\ (\ref{diagcurrentresult}) to analyze the strength of the photon-assisted sidebands more systematically. In the presence of the high-frequency radiation, the electron and hole tunneling rates split into photon-assisted sidebands, see Eq.\ (\ref{GammaSideBands}). To understand the pattern of sidebands in the differential conductance, we assume that the broadening $\Gamma$ is small compared to the photon energy $\Omega$. Then, we can write the total (electron and hole) tunneling rate in Eq.\ (\ref{GammaSideBands}) as $\Gamma(\omega)= [\Gamma_{e,0} + \delta\Gamma_e(\omega)]+[\Gamma_{h,0} + \delta\Gamma_h(\omega)]$. Here, $\Gamma_{e,0}$ and $\Gamma_{h,0}$ denote the contributions of all nonresonant sidebands which are independent of $\omega$ to leading order, while $\delta\Gamma_e(\omega)$ and $\delta\Gamma_h(\omega)$ are the $\omega$-dependent contributions of the resonant sidebands. The distribution of weight over sidebands implies that with one exception discussed below, we can typically neglect the contribution of $\delta\Gamma_e(\omega)$ and $\delta\Gamma_h(\omega)$ to the broadening in the denominator of Eq.\ (\ref{diagcurrentresult}). Similarly, we need to retain the contribution of a resonant sideband in the numerator to obtain a nonzero contribution to the differential conductance. (Recall that the Fermi functions as well as $\Gamma_{e,0}$ and $\Gamma_{h,0}$ are essentially independent of bias voltage.) With these considerations, we obtain the estimate
\begin{equation}
   \left. \frac{\mathrm{d} I}{\mathrm{d} V}\right|_{\mathrm{peak}} \sim \frac{2e^2}{h}\frac{\Gamma_{e,0}\delta\Gamma_{h,{\textrm{thres}}} + \Gamma_{h,0}\delta\Gamma_{e,{\textrm{thres}}}}{(\Gamma_{e,0}+ \Gamma_{h,0})^2},
\label{estside}
\end{equation}
where $\delta\Gamma_{e/h,{\textrm{thres}}}$ is given by evaluating $\delta\Gamma_{e/h}(\omega)$ within $\Gamma_{e,0}+ \Gamma_{h,0}$ of the resonance.

When $|u|^2 = |v|^2$, we have, in magnitude, $\Gamma_{e,0}\approx \Gamma_{h,0}$ and $\delta\Gamma_{e,{\textrm{thres}}}\approx \delta\Gamma_{h,{\textrm{thres}}}$. We then expect electron and hole sidebands to have comparable strengths and we find a peak conductance of order 
\begin{equation}
   \left. \frac{\mathrm{d} I}{\mathrm{d} V}\right|_{\mathrm{peak}} \sim \frac{2e^2}{h}\frac{\delta\Gamma_{e/h,{\textrm{thres}}} }{\Gamma_{e,0}+ \Gamma_{h,0}}.
\end{equation}
Here, we assume for simplicity that the electron and hole sidebands are not overlapping when writing the numerator. Next consider asymmetric electron and hole wavefunctions, say $|u|^2 \ll |v|^2$. Then, we have $\Gamma_{e,0} \ll \Gamma_{h,0}$, and Eq.\ (\ref{estside}) reduces to 
\begin{equation}
   \left. \frac{\mathrm{d} I}{\mathrm{d} V}\right|_{\mathrm{peak}} \sim \frac{2e^2}{h} \left[\frac{\Gamma_{e,0}\delta\Gamma_{h,{\textrm{thres}}} }{\Gamma^2_{h,0}}  + \frac{\delta\Gamma_{e,{\textrm{thres}}}}{\Gamma_{h,0}} \right] .
\end{equation}
At first sight, the first term in the square brackets is suppressed because of the additional factor $\Gamma_{e,0} / \Gamma_{h,0}$. However, the asymmetry between the electron and hole wavefunctions also implies $\delta\Gamma_{e,{\textrm{thres}}}\ll \delta\Gamma_{h,{\textrm{thres}}}$, so that the two terms in the square brackets are of the same order as  they stand. However, the first term is indeed suppressed since it is here where we should remember that the denominator also includes the resonant contributions. For $|u|^2 \ll |v|^2$, these are dominated by $\delta\Gamma_h(\omega)$. This contribution strongly counteracts and thus suppresses the sidebands of the numerator due to $\delta\Gamma_h$. We can then indeed neglect the first term in square brackets and obtain
\begin{equation}
   \left. \frac{\mathrm{d} I}{\mathrm{d} V}\right|_{\mathrm{peak}} \sim \frac{2e^2}{h} \frac{\delta\Gamma_{e,{\textrm{thres}}}}{\Gamma_{h,0}}.
\end{equation}
This explains why hole sidebands are suppressed relative to electron sidebands and thus the appearance of the $Y$-shaped pattern at negative bias voltages as well as the appearance of only a single (electron) set of sidebands at positive bias voltages. Similarly, when $|v|^2 \ll |u|^2$, we have $\Gamma_{h,0} \ll \Gamma_{e,0}$, and we find
\begin{equation}
   \left. \frac{\mathrm{d} I}{\mathrm{d} V}\right|_{\mathrm{peak}} \sim  \frac{2e^2}{h} \frac{\delta\Gamma_{h,{\textrm{thres}}}}{\Gamma_{e,0}},
\end{equation}
so that hole sidebands are dominant. We finally note that these results imply that the sidebands reduce in strength as electron and hole wavefunction become more asymmetric, in agreement with Fig.\ \ref{Fig2_Ac}.
 
Equation (\ref{diagcurrentresult}) also includes the effects of the real part $\Lambda$ of the self energy. It is interesting to note that in the absence of the $ac$ field, the real part does not contribute. Indeed, without $ac$ field, the self energy is either purely real or purely imaginary. Current only flows when both imaginary parts $\Gamma_e$ and $\Gamma_h$  are nonzero, and consequently, $\Lambda(\omega)$ does not contribute. The situation is different in the presence of the $ac$ field, since now the imaginary parts must only be nonzero when absorbing or emitting certain numbers of photons. Contributions to the self energy when absorbing or emitting a different number of photons can still be real and contribute to the resonance denominator in the expression for the current.

Our calculation assumes that we can retain only the contribution of the YSR bound state to the substrate Green function. This requires that the tip-induced broadening of the YSR state remains small compared to the superconducting gap. The characteristic magnitude of the tip density of states is given by the normal-state density of states $\nu_0$ and the YSR wavefunction at the tip position is of order $|u|^2,|v|^2\sim \nu_0\Delta$ \cite{Pientka2013}. This yields the estimate $ |t|^2\nu_0^2\Delta$ for the broadening of the YSR state. Our approximation for the substrate Green function is thus accurate as long as $\nu_0|t|\ll 1$. In view of the normal-state tunneling conductance of the junction, $G_T = (2e^2/h) 4\pi^2 (\nu_0|t|)^2$, this is equivalent to the condition $G_T \ll 2e^2/h$.

\subsection{Exact treatment}
\label{sec:exact}

\subsubsection{Derivation}
\label{sec:calc_ex}

We now consider the exact self energy 
\begin{eqnarray}
 &&\Sigma_R(\tau,\tau^\prime) = |t|^2 \sum_{n, m} J_n({eV_{\mathrm {HF}}}/{\Omega}) J_m({eV_{\mathrm {HF}}}/{\Omega}) 
 \nonumber \\
 &&\,\,\,\,\,\,\times e^{-i(eV+n\Omega) \tau \tau_z}  g_L(\tau-\tau^\prime) e^{i(eV+m\Omega)\tau^\prime\tau_z}.
\end{eqnarray}
including the nondiagonal contribution. In frequency representation defined through
\begin{eqnarray}
   \Sigma_R(\tau,\tau^\prime) = \int \frac{\mathrm{d}\omega}{2\pi} \frac{\mathrm{d}\omega^\prime}
   {2\pi} e^{-i\omega\tau + i\omega^\prime\tau^\prime} \Sigma_R(\omega,\omega^\prime) ,
\end{eqnarray}
this becomes
\begin{eqnarray}
&&\Sigma_R(\omega,\omega^\prime) = |t|^2 \sum_{n, m} J_n({eV_{\mathrm {HF}}}/{\Omega}) J_m({eV_{\mathrm {HF}}}/{\Omega}) \,\,\,\,\,\,\,\,\,\,\,\,\,\,\,\,\,\,\,\,\,\,\,\,
 \nonumber \\
 &&\times 2\pi \delta(\omega-\omega^\prime-(n-m)\Omega\tau_z)   g_L(\omega-(eV+n\Omega)\tau_z) .
\end{eqnarray}
As this is nonzero only when the frequency arguments $\omega$ and $\omega^\prime$ differ by multiples of $\Omega$, we can write
\begin{equation}
   \Sigma_R(\omega,\omega^\prime) = \sum_m 2\pi\delta(\omega-\omega^\prime - m\Omega)     
        \Sigma_m(\omega^\prime)
        \label{Sdelta}
\end{equation}
with 
\begin{eqnarray}
  \Sigma_m(\omega) = \sum_n J_n 
 \left[  \begin{array}{cc}   J_{n+m} g_L(\omega_{-,n}) & 0 \\ 0 & J_{n-m} 
   g_L(\omega_{+,n}) \end{array}\right]. \,\,\,\,\,\,\,\, 
   \label{Sigmamexp}
\end{eqnarray}
Here, we temporarily suppressed the arguments of the Bessel functions and introduced $\omega_{\pm,n} = \omega \pm(eV+n\Omega)$ for compactness. We also note that the self energy satisfies the relation
\begin{equation}
  \Sigma_{-m}(\omega + m\Omega) = \Sigma_m(\omega),
  \label{Sigmamminusm}
\end{equation}
which is readily confirmed using the explicit expression (\ref{Sigmamexp}).

Iteration of the Dyson equation $G_R = g_R +  g_R \Sigma_{R}  G_R$ implies that the Green function $G_R(\omega,\omega^\prime)$ is also nonzero only when its frequency arguments $\omega$ and $\omega^\prime$ differ by multiples of $\Omega$. Thus, we define
\begin{equation}
   G_R(\omega,\omega^\prime) = \sum_m 2\pi\delta(\omega-\omega^\prime - m\Omega)     
        G_m(\omega^\prime)
\label{Gdelta}
\end{equation}
with 
\begin{equation}
   G_R(\tau,\tau^\prime) = \sum_n \int \frac{\mathrm{d}\omega}{2\pi} e^{-i\omega(\tau-\tau^\prime)-
   in\Omega\tau} G_n(\omega).
\end{equation}
Inserting Eqs.\ (\ref{Sdelta}) and (\ref{Gdelta}) into the Dyson equation, we find
\begin{eqnarray}
   &&G_n(\omega) = g_R(\omega)\delta_{n,0} \nonumber\\
   &&\,\,\,\,\,\,\,\,+ \sum_m g_R(\omega+n\Omega ) \Sigma_{n-m}(\omega+m\Omega) G_m(\omega),
   \label{GreenFinal}
\end{eqnarray}
which provides a set of linear equations to compute the $G_n(\omega)$. 

Writing the current in Eq.\ (\ref{basiccurrent}) using Eqs.\ (\ref{Sdelta}) and (\ref{Gdelta}) and focusing on the $dc$ contribution, we find
\begin{eqnarray}
  && I_{\mathrm {dc}} = e\int \frac{\mathrm{d}\omega}{2\pi} \sum_n \mathrm{Tr} \Big\{ \tau_z  \nonumber\\
  && \times \left[ G_n^<(\omega)\Sigma^a_{-n}(\omega+n\Omega) + G_n^r(\omega)\Sigma^<_{-n}(\omega+n\Omega) \right. \nonumber\\
 && \left. -\Sigma_n^<(\omega)G^a_{-n}(\omega+n\Omega)-\Sigma_n^r(\omega)G^<_{-n}(\omega+n\Omega)\right]\Big\}.
\end{eqnarray}
This can be made more compact by using Eq.\ (\ref{Sigmamminusm}),
\begin{eqnarray}
  && I_{\mathrm {dc}} = e\int \frac{\mathrm{d}\omega}{2\pi} \sum_n \mathrm{Tr} \Big\{ \tau_z 
  \left[ G_n^<(\omega)\Sigma^a_{n}(\omega) + G_n^r(\omega)\Sigma^<_{n}(\omega)\right. \nonumber\\
  &&  \,\,\,\,\,\,\,\,\,\,\,\,\,\,\,\,\,\,\,   \left.   -\Sigma_n^<(\omega)G^a_{n}(\omega)-\Sigma_n^r(\omega)G^<_{n}(\omega)\right]\Big\}.
  \label{ExactCurrentFinal}
\end{eqnarray}
Together with the expressions (\ref{Sigmamexp}) and (\ref{GreenFinal}) for the self energy and the Green function, respectively, this constitutes our final result. 

\begin{figure}[t]
\centering
\includegraphics[width=0.8\columnwidth]{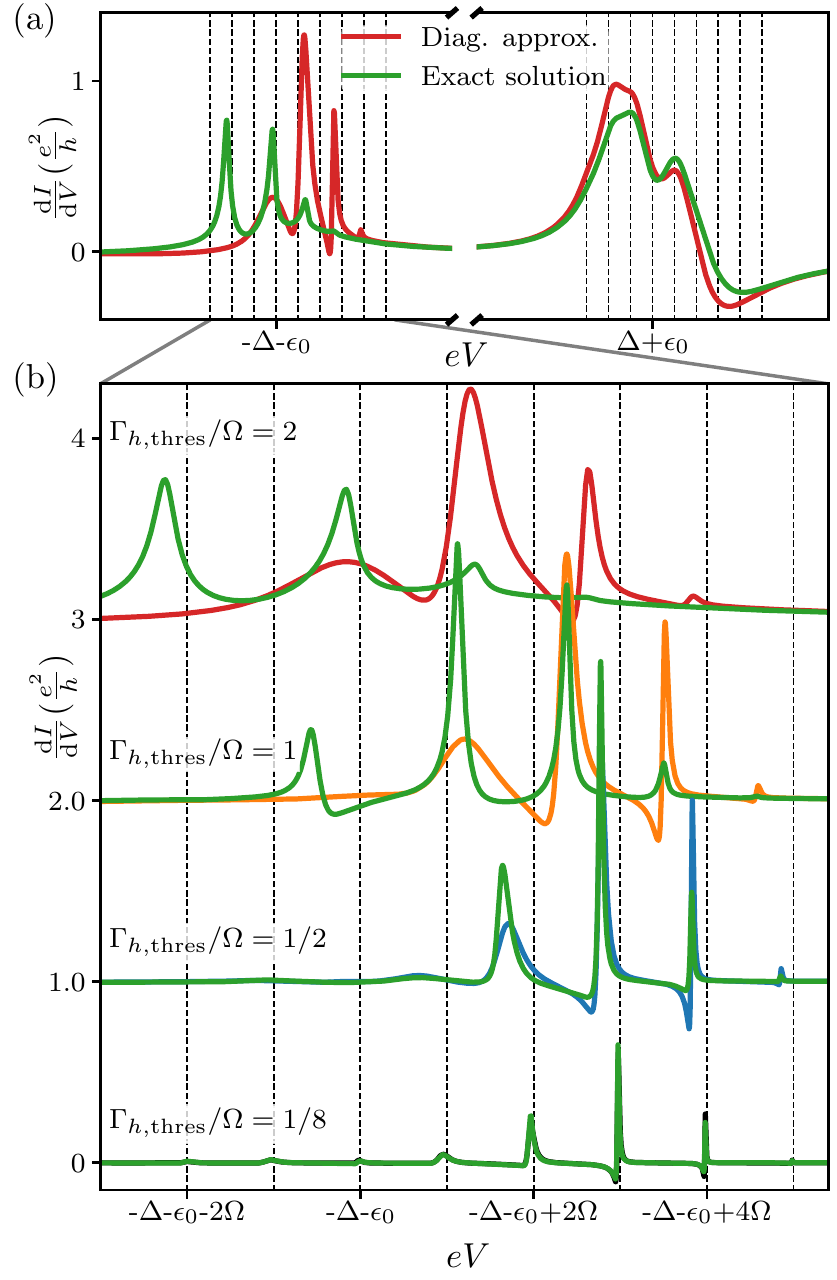}
\caption{Comparison between diagonal approximation (different colors) and the exact solution (green) for $\epsilon_0/\Delta = 0.4$, $\Omega/\Delta = 0.025$, $eV_\mathrm{HF} = 2\Omega$, and $u^2=v^2/9$. (a) $\mathrm{d}I/\mathrm{d}V$ at negative and positive voltages at strong tip-sample tunneling ($\Gamma_{h,\mathrm{thres}} = 2\Omega$). The resonances at negative bias voltages are shifted relative to the diagonal approximation, while the differences are merely quantitative at positive biases. (b) Closeup of threshold region at negative bias voltages for increasing tip-sample tunneling $\Gamma_{h,\mathrm{thres}}$ as indicated in the figure (from bottom to top; offset for clarity). The differences between exact solution (green) and diagonal approximation become substantial once $\Gamma_{h,\mathrm{thres}}$ becomes comparable to the photon energy $\Omega$. Dashed lines indicate multiples of the photon energy $\Omega$. 
\label{Fig6_Ac}}
\end{figure}

\subsubsection{Results}

We solve Eq.\ (\ref{GreenFinal}) numerically by truncating the system of equations at a sufficiently high $|n|\gg V_\mathrm{HF}/\Omega$ and compute the current from Eq.\ (\ref{ExactCurrentFinal}). Due to the terms of the self energy which are offdiagonal  in frequency,  
the exact solution is sensitive to Green functions which are evaluated at frequencies shifted by integer multiples of the photon energy. This suggests that the exact solution deviates from the diagonal approximation when the tunneling-induced broadening of the Green functions becomes of the order of or larger than the photon energy. Conversely, the diagonal approximation is expected to be accurate in the limit  of small broadening and well-resolved photon sidebands. 

Figure \ref{Fig6_Ac} compares representative numerical results obtained in the diagonal approximation and the numerically exact solution. The results consider the parameter regime $u^2 = v^2/9$ where the Y shape appears at negative voltages. The choice of  $eV_\mathrm{HF}/\Omega = 2$ implies that the resonances at negative voltages are associated with the lower part (stem) of the Y shape. Panel (a) shows both negative and positive voltages for strong tunneling-induced broadening. At positive biases, we find that the sidebands are no longer well resolved due to the broadening and the differences between the diagonal approximation and exact result are largely quantitative. The self energy already present in the diagonal approximation, including the hole contribution to the broadening, dominates over additional contributions in the exact solution. In contrast, we find distinct differences at negative voltages. Here, the hole contribution to the broadening is still suppressed around the threshold voltage along the stem of the Y shape and the resonances remain well resolved. One then observes that the sidebands are distinctly shifted to higher bias voltages in the exact solution, while the width of the resonances remains essentially unchanged, i.e., the dominant effect is associated with the real part of the self energy.  

Panel (b) explores the dependence on the strength of tip-substrate tunneling, focusing on the region of negative voltages. We quantify the four different strengths of tip-substrate tunneling by the threshold value for the hole tunneling rate
\begin{equation}
 \Gamma_{h,\mathrm{thres}} = \frac{1}{2}\left(\gamma_h^2 \Delta \right)^{1/3},
\end{equation}
as evaluated for the regime of dominant hole tunneling. For weak tip-substrate tunneling, $\Gamma_{h,\mathrm{thres}}/\Omega = 1/8$, the broadening is small compared to the photon energy and in agreeement with expectations, the diagonal approximation is essentially identical to the exact solution. For $\Gamma_{h,\mathrm{thres}}/\Omega = 1/2$, quantitative differences such as modified peak heights begin to appear, but the peak positions still remain identical. The differences become more pronounced for $\Gamma_{h,\mathrm{thres}}/\Omega = 1$ and $\Gamma_{h,\mathrm{thres}}/\Omega = 2$, where we observe substantial shifts of the peaks to higher bias voltages. Also note that the resonance width grows with increasing $\Gamma_{h,\mathrm{thres}}/\Omega$ as expected. These results show that the diagonal approximation is accurate in the regime of well-resolved sidebands. 

We finally point out that the diagonal approximation is exact for a normal-state tip with a constant density of states $\nu_0$. In this case,  the retarded and advanced self energies in Eq.\ (\ref{selfenergy}) are purely imaginary,
\begin{equation}
   \Sigma^{r,a}_R(\tau,\tau^\prime) = \mp i\pi |t|^2 \nu_0 \delta(\tau-\tau^\prime),
\end{equation}
and independent of the $ac$ field. This makes also the retarded and advanced substrate Green functions independent of the $ac$ field, so that $\Sigma_n^{r,a}$ is nonzero for $n=0$ only,
\begin{eqnarray}
   \Sigma_{n=0}^{r,a}(\omega) &=&  \mp i\pi |t|^2 \nu_0  .
\end{eqnarray}
Then, $G^{r,a}_{n}$ is nonzero for $n=0$ only and only the $n=0$ term contributes to the $dc$ current in Eq.\ (\ref{ExactCurrentFinal}). Moreover, one readily ascertains from the Dyson equation (\ref{GreenFinal}) and the Langreth rules that the $n=0$ components satisfy a closed set of equations which is just the set of equations which leads to the diagonal approximation. 

\begin{figure}[t]
\includegraphics[width=.8\columnwidth]{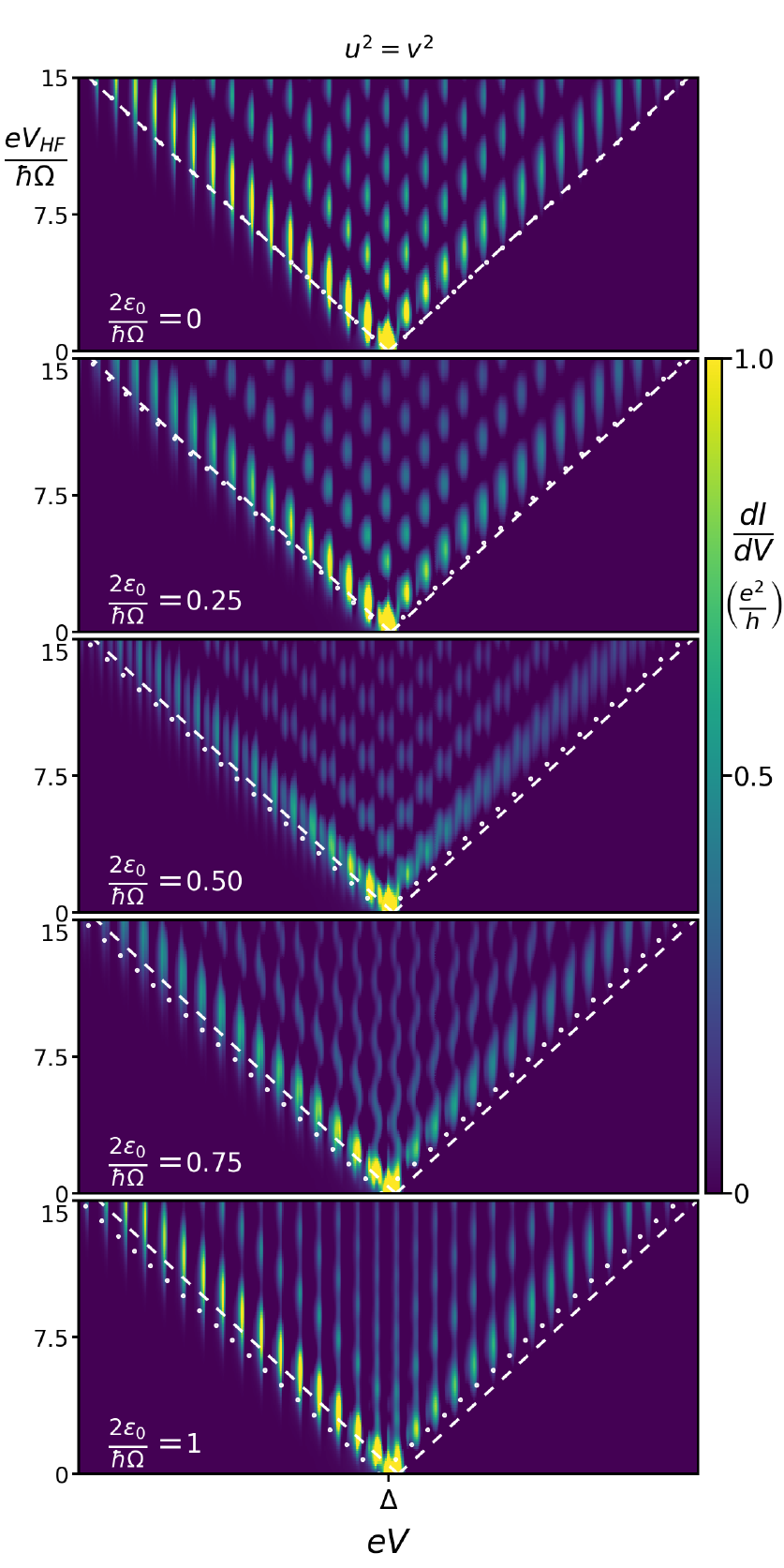}
\caption{Differential conductance (color scale) as a function of bias voltage $eV$ and amplitude $eV_{\rm HF}$ of the high-frequency radiation for tunneling into a YSR state with equal electron and hole wavefunctions, $|u|^2=|v|^2$ and small YSR energies $\epsilon_0$ increasing from zero to $\hbar\Omega/2$ from top to bottom as indicated in the panels. The regions with electron and hole sidebands are indicated by white dashed and dotted lines, respectively. The five panels show clearly that a nonzero energy of the YSR state generates a splitting of the photon-assisted sidebands which appears with high multiplicity throughout the V-shaped region. This provides the basis for a high-resolution measurement of the energy of the subgap state, which can be used to identify YSR (or Andreev) bound states with near-zero energy $\epsilon_0$ with high resolution, and thereby distinguish them from Majorana bound states. Parameters: $\Omega/\Delta = 0.05$, $\nu_0|t| = 0.04$.
\label{Fig7_Ac}}
\end{figure}

\section{Majorana bound states}
\label{sec:Majorana}

\begin{figure*}[t]
\centering
\includegraphics[width=1.0\textwidth]{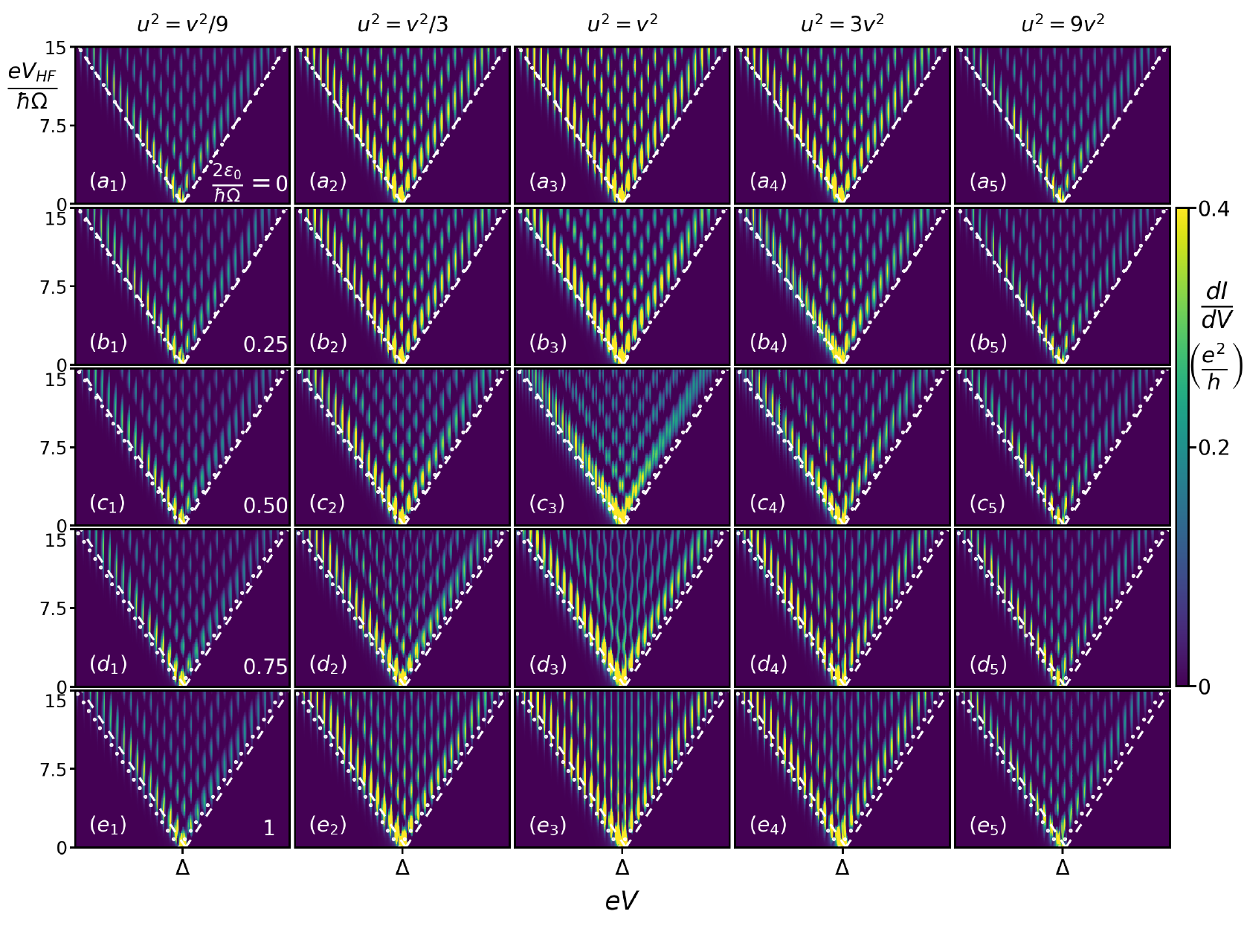}
\caption{Differential conductance (color scale) as a function of bias voltage $eV$ and amplitude $eV_{\rm HF}$ of the high-frequency radiation for tunneling into a YSR state with various ratios of electron and hole wavefunctions and small YSR energies $\epsilon_0$ increasing from zero to $\hbar\Omega/2$ from top to bottom, as indicated in the panels. The regions with electron and hole sidebands are outlined by white dashed and dotted lines, respectively. The splitting of the sidebands due to a small nonzero energy $\epsilon_0$ appearing for equal electron and hole wavefunctions (central column of panels) are less pronounced for asymmetric electron and hole wavefunctions. For YSR states, the ratio of electron and hole wavefunctions typically varies as a function of position. In STM experiments, one can therefore generically choose a tip position for which electron and hole wavefunctions have similar magnitude. Parameters: $\Omega/\Delta = 0.05$, $\nu_0|t| = 0.04$.
\label{Fig8_Ac}}
\end{figure*}

\subsection{Basic results}

Our considerations for YSR states apply to photon-assisted tunneling into Majorana bound states with only minor modifications. First, Majorana bound states have zero energy so that we set $\epsilon_0=0$. Second, their electron and hole wavefunctions are equal in magnitude, satisfying $u=v^*$ for spinless fermions (and corresponding expressions for spinful electrons in a four-component Nambu formalism). Finally, an isolated Majorana bound state is a solution of the particle-hole symmetric Bogoliubov-deGennes equation which doubles the degrees of freedom. Due to this doubling of degrees of freedom, the expression for the current must be multiplied by a factor of $1/2$ relative to the case of a YSR state. We note that here, we focus on tunneling into Majorana bound states in a grounded superconductor. A recent experiment \cite{Zanten2019} has studied the effects of photon-assisted tunneling on the charge stability diagram of two coupled Majorana nanowires with floating superconductors subject to charging energies.

First consider photon-assisted resonant Andreev processes into Majorana bound states from a normal-metal tip. Using these translation rules, we obtain corresponding results directly from the results for YSR states given in Sec.\ \ref{sec:normalmetal}. The equal magnitude of electron and hole wavefunctions makes the peak conductance universal and equal to $2e^2/h$ for Majorana bound states \cite{Law2009,Flensberg2010}. This corresponds to half of the maximal peak conductance of a zero-energy YSR state, reflecting that Majorana bound states are effectively only half of an ordinary subgap state. 

In the presence of high-frequency radiation, there are photon-assisted sidebands and one readily obtains from Eq.\ (\ref{SupTipDiag}) that 
\begin{eqnarray}
   \frac{\mathrm {d}I}{\mathrm {d}V} = \frac{2e^2}{h}\sum_n J_n^2({eV_{\mathrm {HF}}}/
  {\hbar\Omega})
   \frac{\gamma^2}{(eV +n\hbar\Omega)^2 
  +{\gamma^2}},\,\,\,\,\,\,
\end{eqnarray} 
where we introduced $\gamma=\gamma_e=\gamma_h$. Thus, the familiar Majorana zero-bias peak of height $2e^2/h$ splits into photon sidebands with a sideband spacing in bias voltage of $\hbar\Omega/e$. As for YSR states, this can be traced back to the existence of separate threshold conditions for electrons and holes. For Majorana bound states, these two sets of conditions coincide by particle-hole symmetry, leading to a sideband spacing of $\hbar\Omega/e$ seemingly indicating single-electron tunneling despite the underlying resonant Andreev process.  

For a superconducting tip, we focus on the limit of well-resolved sidebands where the diagonal approximation (\ref{diagcurrentresult}) is accurate and obtain
\begin{eqnarray}
  && I = e\, \mathrm{sgn}(V) \int \frac{\mathrm{d}\omega}{2\pi} \sum_{n,m}J_n^2({eV_{\mathrm {HF}}}/
  {\Omega})J_m^2({eV_{\mathrm {HF}}}/{\Omega}) \nonumber\\
  && \,\,\,\,\,\,\,\,\,\times \frac{\Gamma_e(\omega\!-\!(eV \!+\! n\Omega))\Gamma_h(\omega\!+\!(eV \!+\! m\Omega))}{[\omega-\epsilon_0-\Lambda(\omega)]^2 + 
    \frac{1}{4}\Gamma^2(\omega)},  
   \label{diagcurrentresultMajorana}
\end{eqnarray}
where $\Gamma_e(\omega)$ and $\Gamma_h(\omega)$ are now evaluated with $|u|^2=|v|^2$ and thus equal. Up to an overall scale factor of $1/2$, the result is identical to that for a YSR state with $\epsilon_0=0$ and $|u|^2=|v|^2$ as shown in the top panel in Fig.\ \ref{Fig7_Ac}. For $\epsilon_0=0$, the V shapes for the electron and hole conditions coincide and are centered on $eV=\pm\Delta$. This also implies that similar to the case of a normal-state tip, there is only one set of sidebands with spacing $\hbar\Omega$ which is enhanced by the fact that electron and hole resonances coincide. While the pattern of resonances at $eV=\pm\Delta$ is necessarily symmetric with respect to a change of sign of the bias voltage, the individual V shapes are asymmetric about $eV=\Delta$ (or, analogously, $eV=-\Delta$). This is a consequence of the fact that the broadening is smaller on the small-bias side of the V shape, leading to sharper features and a larger differential conductance (see Fig.\ \ref{Fig7_Ac}; this is not properly reflected by the color scale in Fig.\ \ref{Fig8_Ac} due to saturation effects).  

\subsection{Majorana vs.\ YSR states}

It is frequently a challenge to distinguish zero-energy Majorana bound states from other low-energy subgap states. Moreover, in many experiments, putative Majorana states might be accompanied by close-lying YSR states \cite{NadjPerge2014,Ruby2015, Feldman2017, Kim2019, Schneider2020}. 
Our results on YSR and Majorana bound states imply that photon-assisted tunneling provides a high-resolution method to determine the energy of subgap states. In principle, superconducting tips are preferable over normal-metal tips because the gap suppresses thermal excitations and the strongly peaked BCS density of states allows for high energy resolution. At the same time, ${\mathrm d}I/{\mathrm d}V$ peaks due to subgap states with a zero or small energy $\epsilon_0$ no longer appear as (near) zero-bias peaks, but rather at $eV=\Delta+\epsilon_0$ (since the tunneling electron leaves behind a quasiparticle in the tip) \cite{NadjPerge2014,Ruby2015, Feldman2017}. Thus, the small energy $\epsilon_0$ of the bound states is effectively extracted as a difference of two much larger energies. In particular, this implies that inaccuracies in the determination of the tip gap carry over fully into the accuracy with which the bound-state energy can be determined. 

The existence of independent thresholds for electron and hole tunneling in photon-assisted resonant Andreev reflections provides a method to extract the bound-state energy from a line splitting which appears directly in the measured tunneling spectra \cite{Peters2020}. Moreover, this line splitting appears with a high multiplicity throughout the V-shaped region within which one observes thresholds for photon-assisted tunneling. To illustrate this, consider first resonant Andreev reflections into a YSR state with equal electron and hole wavefunctions and a small energy $\epsilon_0$, as shown in Fig.\ \ref{Fig7_Ac}. Up to an overall prefactor of 1/2, the panel for $\epsilon_0=0$ is identical to the result for a Majorana bound state. One observes that even a small $\epsilon_0$ which is just a fraction of the photon energy $\hbar\Omega$ leads to a splitting of the sidebands and can thus be accurately detected. This is most evident for $\epsilon_0=\hbar\Omega/4$, making experiments with variable photon energies particularly advantageous.

In addition to the line splitting, there is also a characteristic change in the dependence of the sideband strengths as a function of the amplitude $V_{\mathrm{HF}}$ of the high-frequency radiation. As seen in Fig.\ \ref{Fig7_Ac} and  Fig.\ \ref{Fig8_Ac}, the sideband strengths exhibit repeated zeros as a function of $V_{\mathrm{HF}}$. The zeros originate from the oscillations of the Bessel functions in Eq.\ (\ref{diagcurrentresult}) [see also Eq.\ (\ref{Besselexp})]. Physically, these can be considered a result of interference between various sequences of emissions and absorptions of `photons’ contributing to a sideband. (Notice that the sideband strength is nonperturbative in $V_{\mathrm{HF}}$ and emerges from processes of all orders when viewed from the point of view of perturbation theory.) Different sidebands are controlled by Bessel functions of different order, and the corresponding phase shift leads to a shift in the locations of the zeros between neighboring sidebands.

The behavior of the zeros in Fig.\ \ref{Fig7_Ac} then emerges as follows. The separate thresholds for electron and hole tunneling coincide for $\epsilon_0=0$, but move apart when $\epsilon_0$ becomes nonzero. When $\epsilon_0=0$, a particular sideband `combines’ electron and hole sidebands described by Bessel functions of the same order, and the zeros of the Bessel functions are preserved. For $2\epsilon_0=\hbar\Omega$, the electron and hole sidebands are described by Bessel functions of neighboring orders. Their zeros no longer coincide and thus the zeros in the observed sideband strengths disappear. This allows one to distinguish true zero-energy states from situations with nonzero $\epsilon_0$ in which electron and hole sidebands coincide because $2\epsilon_0$ and $\hbar\Omega$ are commensurate.

Corresponding results with unequal electron and hole wavefunctions are shown in Fig.\ \ref{Fig8_Ac}. Clearly, the splitting due to a small $\epsilon_0$ is most pronounced for equal electron and hole wavefunctions, for which sidebands emerging from electron and hole sidebands are both equally prominent, cp.\ the discussion in Sec.\ \ref{sec:suptip}. For a YSR state, the ratio of electron and hole wavefunctions varies as a function of position. In STM experiments, one should thus choose a tip position where electron and hole wavefunctions are equal to optimize sensitivity. Finally, notice that the modulations of the sideband strength as a function of $V_{\mathrm{HF}}$ reemerge even for $2\epsilon_0=\hbar\Omega$ once electron and hole wavefunctions are sufficiently different. In this case, the electron and hole thresholds contribute with different strengths, and the sidebands are dominated by one or the other.

\section{Conclusions}
\label{sec:conclusions}

We have developed a theory for photon-assisted resonant Andreev tunneling into subgap states in superconductors. Our results are in excellent agreement with recent STM measurements on YSR states \cite{Peters2020}, fully reproducing the observed patterns of sidebands which differ markedly from predictions of a simple Tien-Gordon-like theory. 

A central aspect of the theory are independent sideband conditions for the electron and hole tunneling processes. This leads to two sets of sidebands whose relative shift in bias voltage depends on the ratio of the energy of the subgap state and the photon energy. As an interesting consequence, this provides a sensitive technique to measure near-zero energies of subgap states which can be instrumental in distinguishing conventional subgap states from Majorana bound states. Simultaneous visibility of the two sets of sidebands is optimal when electron and hole wavefunctions are of similar magnitude. For YSR states, the ratio of electron and hole wavefunctions typically varies widely with lateral position \cite{Yazdani1997, Menard2015, Ruby2016, Choi2017}. This can be exploited in STM experiments by choosing an appropriate lateral position of the STM tip for optimal resolution. The absence of spatial resolution may make this technique less flexible in transport experiments using gate defined tunnel junctions. 

The observability of the photon-assisted sidebands of resonant Andreev reflections is constrained by two requirements. On the one hand, tunneling should be sufficiently weak for the tunneling-induced broadening to be small compared to the sideband spacing so that the sidebands are well resolved. At the same time, the underlying resonant Andreev reflections require tunneling to be fast compared to inelastic relaxation processes. The latter provide competing channels which transfer electrons into the quasiparticle continuum of the substrate via the subgap state and which transfer only a single electron between tip and substrate. A recent STM experiment using a photon frequency of 40 GHz shows that these conditions on the junction conductance can be simultaneously satisfied at a temperature of order 1K. Since inelastic excitations have an activated temperature dependence, their rate drops rapidly as temperature is lowered. This implies that even though the broadening of the sidebands is independent of temperature, the attainable resolution of small subgap energies improves rapidly at lower temperatures. 

In view of recent experiments, we have focused on YSR states throughout the paper. However, we emphasize that our theoretical approach is in no way specific to YSR states and applies equally well to other subgap states. Consequently, photon-assisted tunneling could also contribute to distinguishing Andreev from Majorana bound states. 

At present, our theory assumes a single subgap state. However, magnetic impurities frequently induce multiple subgap states within the superconducting gap \cite{Ji2008, Ruby2016, Choi2017, Hatter2015}. It would thus be interesting to extend the theory to include several subgap states where one would expect additional spectroscopic features to arise when the photon energy becomes comparable to the level spacing between subgap states. 

\begin{acknowledgments}
We gratefully acknowledge funding by Deutsche Forschungsgemeinschaft through CRCs 183 and 910 (FvO), by the European Research Council under Consolidator Grant {\em NanoSpin} (KJF), and the Danish National Research Foundation as well as  the Independent Research Fund Denmark | Natural Sciences (KF). One of us (KF) is also grateful for the hospitality of the Dahlem Center for Complex Quantum Systems in the context of a Mercator Professorship funded by the Deutsche Forschungsgemeinschaft.  
\end{acknowledgments}

%


\begin{thebibliography}{61}
\bibitem{foot10} We refer generically to high-frequency radiation to encompass, e.g., radiofrequency or microwave radiation.
\end{thebibliography}

\appendix

\section{Tip Green function}
\label{app:gL}

In this appendix, we briefly review the derivation of the tip Green function. The Nambu Green function $g$ of a BCS superconductor (in the absence of tunneling or a magnetic impurity) takes the form
\begin{equation}
    g_L({\bf k},\omega) = [\omega - \xi_{\bf k} \tau_z - \Delta \tau_x]^{-1},
\end{equation} 
where $\tau_x$ and $\tau_z$ denote Pauli matrices in Nambu space. Performing the matrix inversion and computing the corresponding local Green function at the tip position yields
\begin{equation}
    g_L(\omega) = \frac{1}{V}\sum_k g({\bf k},\omega) = \nu_0 \int {\mathrm d}{\bf k} 
    \frac{\omega + \xi_{\bf k} + \Delta \tau_x}{\omega^2 - \xi_{\bf k}^2 - \Delta^2}. 
\end{equation}
Performing the integral gives the result
\begin{equation}
   g_L(\omega) = - \frac{\pi \nu_0 (\omega + \Delta \tau_x)}{\sqrt{\Delta^2-\omega^2}}.
\end{equation}
This can be used to find the retarded and advanced as well as lesser Green functions which are used throughout the main text.

The retarded and advanced Green functions are purely real at frequencies below the gap, $|\omega|<\Delta$, where one finds
\begin{equation}
   g^{r/a}_L(\omega) = - \frac{\pi \nu_0 (\omega + \Delta \tau_x)}{\sqrt{\Delta^2-\omega^2}},
\end{equation}
and purely imaginary at frequencies above the gap, $|\omega|>\Delta$,
\begin{equation}
   g^{r/a}_L(\omega) = \mp \frac{ i \pi \nu_0 (\omega + \Delta \tau_x)}{\sqrt{\omega^2-\Delta^2}}        
    \,\mathrm{sgn} (\omega).
\end{equation} 
To derive the lesser Green function, we use the relation
\begin{equation}
  g_L^< (\omega) = - n_F(\omega) [g^{r}_L(\omega) -g^{a}_L(\omega)]
  \label{gL_app}
\end{equation}
and obtain
\begin{equation}
  g_L^< (\omega) = 2\pi  i n_F(\omega) \frac{\nu_0 (\omega + \Delta \tau_x)}{\sqrt{\omega^2-\Delta^2}}
   \theta(|\omega| - \Delta) \mathrm{sgn} (\omega) ,
\end{equation}
where $\theta(x)$ denotes the Heaviside function.

Within our calculation, neglecting Andreev reflections in the tip is equivalent to dropping the off-diagonal contributions to the tip Green function. In this approximation, $g_L(\omega)$ becomes proportional to the unit matrix in Nambu space and we find 
\begin{equation}
   g^{r/a}_L(\omega) \simeq \left\{  \begin{array}{ccc}  
   - \pi \nu_0  \frac{ \omega }{\sqrt{\Delta^2-\omega^2}}, & &  |\omega|<\Delta \\
   \mp i \pi \nu_0 \frac{ |\omega| }{\sqrt{\omega^2-\Delta^2}},        
     & & |\omega|>\Delta \end{array} \right. 
     \label{gLra}
\end{equation}
for the retarded and advanced Green functions and 
\begin{equation}
  g_L^< (\omega) \simeq 2\pi  i n_F(\omega) \frac{\nu_0 |\omega|}{\sqrt{\omega^2-\Delta^2}}
  \, \theta(|\omega| - \Delta)
  \label{gLless}
\end{equation}
for the lesser Green function. The above-gap expressions can be expressed compactly in terms of the BCS density of states in Eq.\ (\ref{BCSDOS}).

\section{Substrate Green function}
\label{app:SubG}

This appendix discusses the Green function of the substrate. We first consider the bare substrate Green function at subgap energies. In keeping with our approximation of neglecting the (nonresonant) Andreev reflections in the tip, we retain only the bound state contributions to the substrate Green function which are responsible for the resonant Andreev reflections. For general (spinful) Hamiltonians, one needs to work with Nambu operators which involve electrons and holes of both spins. The resulting Bogoliubov-deGennes equation 
is particle-hole symmetric and subgap bound states will appear in pairs with energies of opposite sign. Correspondingly, in this approach, one finds pairs of YSR states with energies $\pm\epsilon_0$. Such spinful Nambu and Bogoliubov-deGennes descriptions are however redundant in that they double the degrees of freedom. 

In its spinful version, the Bogoliubov-deGennes Hamiltonian of the present problem is block-diagonal, with one subspace spanned by spin-up electrons and spin-down holes, and the other subspace by spin-down electrons and spin-up holes (with the spin-quantization axis taken parallel to the impurity spin). The two subspaces are related by particle-hole symmetry. This has two important consequences. First,
the doubling of degrees of freedom can be avoided by retaining only one of the two subspaces. Second, each subspace hosts one of the partners of each pair of YSR states. Consequently, when retaining only one subspace, there is only one YSR state with Bogoliubov-deGennes wavefunction $\psi^T=(u ,v)$ at the tip position ${\bf R}$. Thus, we find
\begin{equation}
    g_R(\omega) = \psi \frac{1}{\omega-\epsilon_0} \psi^\dagger
\end{equation} 
for the approximate (bare) substrate Green function at subgap energies. (We assume here for simplicity that there is only one pair of YSR states in the spinful formulation.) 

Tunneling introduces a self energy into the denominator of the retarded and advanced Green functions, 
\begin{equation}
    G^{r/a}_R = \psi \frac{1}{ \omega-\epsilon_0 - \tilde \Sigma_R^{0,r/a}} \psi^\dagger
\end{equation} 
with 
\begin{equation}
   \tilde \Sigma_R^{0,r/a} = \psi^\dagger \Sigma_R^{0,r/a}\psi.
\end{equation}
Note that we have written the last two expressions in general operator notation since with $ac$ field, the self energy is generally no longer diagonal in frequency representation.
 
We also review a general relation for the lesser Green function (including the tunneling to the tip). Using the Langreth rules, the Dyson equation for $G_R$ gives
\begin{equation}
   G_R^< = g_R^< + g_G^r \Sigma_R^r G_R^< + g_G^r \Sigma_R^< G_R^a + g_G^< \Sigma_R^a 
     G_R^a,
\end{equation}
which can be readily shown to become 
\begin{equation}
   G_R^< = \frac{1}{1-g_R^r \Sigma_R^r} g_R^< \frac{1}{1-g_R^a \Sigma_R^a} + G_R^r \Sigma_R^< 
    G_R^a.
\end{equation}
The first term on the right-hand side vanishes generally as long as the system was noninteracting in the infinite past. Here, we can also use the explicit expression (\ref{gL_app}), with $L$ replaced by $R$, to write
\begin{eqnarray}
    &&\frac{1}{1-g_R^r \Sigma_R^r} g_R^< \frac{1}{1-g_R^a \Sigma_R^a} \nonumber\\
     && = -n_F(\omega) \frac{1}{1-g_R^r \Sigma_R^r} [g_R^r - g_R^a] \frac{1}{1-g_R^a \Sigma_R^a}.
\end{eqnarray}
Inserting the identity 
\begin{equation}
   g_R^r-g_R^a = -2i\eta g_R^r g_R^a
\end{equation}
with a positive infinitesimal $\eta$ yields
\begin{eqnarray}
   \frac{1}{1-g_R^r \Sigma_R^r} g_R^< \frac{1}{1-g_R^a \Sigma_R^a} = 2i\eta n_F(\omega)  G_R^r  G_R^a = 0. \,\,\,\,\,\,\,\,\,
\end{eqnarray}
Thus, we find the identity
\begin{equation}
     G_R^< = G_R^r \Sigma_R^< G_R^a.
     \label{Glessrlessa}
\end{equation}

\section{Peak differential conductance}
\label{app:dIdV}

In this appendix, we sketch the derivation of the expressions for the differential conductance given in Sec.\ \ref{sec:suptip}. We focus on the case of positive bias voltage $eV=\Delta +\epsilon_0$. The other cases can be obtained in an analogous manner. To start with, the current (without high-frequency radiation) is given by Eq.\ (\ref{diagcurrentresult}), where for $T\ll \delta$ we can set the Fermi functions to zero and one, respectively,
\begin{equation}
 I = 2e\int \frac{\mathrm{d}\omega}{2\pi} 
  \frac{\Gamma_e(\omega - eV )\Gamma_h(\omega + eV )}{[\omega-\epsilon_0]^2 + 
    \frac{1}{4}[\Gamma_e(\omega-eV) + \Gamma_h(\omega+eV)]^2}.
\end{equation}
For $eV \simeq eV_0 = \Delta+\epsilon_0$, $\Gamma_h(\omega+eV)$ is nonsingular, so that we can neglect the bias dependence and set $\omega \simeq \epsilon_0$ due to the resonance denominator. Assuming also that $\epsilon_0$ is small compared to $\Delta$ and large compared to the broadening of the resonance, this yields $\Gamma_{h,\mathrm{thres}}$ as given in Eq.\ (\ref{Gammahthres}). (We note that the assumption of $\epsilon_0\ll\Delta$ is in no way essential and can be easily lifted.) In contrast, the bias dependence cannot be neglected in the electron tunneling rate, since the latter becomes singular at the threshold. Thus, we write $V= V_0 + \delta V$ and obtain 
\begin{equation}
 I \simeq 2e\int \frac{\mathrm{d}\omega}{2\pi} 
  \frac{\Gamma_e(\omega_- -e\delta V )\Gamma_{h,\mathrm{thres}}}{[\omega-\epsilon_0]^2 + 
    \frac{1}{4}[\Gamma_e(\omega_-  - e\delta V) + \Gamma_{h,\mathrm{thres}}]^2}.
\end{equation}
Here, we introduced the shorthand $\omega_-=\omega - eV_0$. Shifting the integration variable, $\omega \to \omega + e\delta V$, this becomes
\begin{equation}
 I \simeq 2e\int \frac{\mathrm{d}\omega}{2\pi} 
  \frac{\Gamma_e(\omega_- )\Gamma_{h,\mathrm{thres}}}{[\omega-\epsilon_0 + e\delta V]^2 + 
    \frac{1}{4}[\Gamma_e(\omega_-) + \Gamma_{h,\mathrm{thres}}]^2}.
\end{equation}
This yields
\begin{equation}
 \left.\frac{\mathrm{d}I}{\mathrm{d}V}\right|_{\mathrm{peak},+} \simeq \int \frac{\mathrm{d}\omega}{2\pi} 
  \frac{4e^2 (\epsilon_0 - \omega) \Gamma_e(\omega_-)\Gamma_{h,\mathrm{thres}}   }{([\omega-\epsilon_0 ]^2 + 
    \frac{1}{4}[\Gamma_e(\omega_-) + \Gamma_{h,\mathrm{thres}}]^2)^2}.
\label{C4}
\end{equation}
for the differential conductance at $V=V_0$. Using $\Gamma_e(\omega) = \gamma_e [\nu(\omega)/\nu_0]$, one readily finds the expression for $\Gamma_e(\omega_-)$ in Eq.\ (\ref{Gammaethres}). Note that due to the $\theta$-function in this expression, the integral in Eq.\ (\ref{C4}) ranges effectively over $\omega$ from $-\infty$ to $\epsilon_0$. 

First consider the situation that the broadening of the resonance denominator is dominated by hole tunneling. Then, the integral is dominated by $\epsilon_0 - \omega \sim \Gamma_{h,\mathrm{thres}}$. Using this in the expression for the electron tunneling rate, we find that hole tunneling dominates provided that $\Gamma_{h,\mathrm{thres}}\gg (\gamma^2_e\Delta)^{1/3}$. The characteristic electron tunneling rate is then given by $\Gamma_{e,\mathrm{thres}}=\gamma_e[\Delta/2\Gamma_{h,\mathrm{thres}}]^{1/2}$. Consequently neglecting the contribution of electron tunneling to the broadening of the denominator, we readily find
\begin{equation}
 \left.\frac{\mathrm{d}I}{\mathrm{d}V}\right|_{\mathrm{peak},+} \simeq \frac{4 e^2}{h}\frac{\Gamma_{e,\mathrm{thres}}}{\Gamma_{h,\mathrm{thres}}} \int_0^{\infty} \frac{  \mathrm{d}x\, x^{1/2}}{(x^2+1/4)^2}.
\end{equation}
Performing the integral yields the final expression
\begin{equation}
 \left.\frac{\mathrm{d}I}{\mathrm{d}V}\right|_{\mathrm{peak},+} \simeq \frac{4\pi e^2}{h}\frac{\Gamma_{e,\mathrm{thres}}}{\Gamma_{h,\mathrm{thres}}} = \frac{8\pi e^2}{h}\frac{\gamma_e \epsilon_0^{3/4}}{\gamma_h^{3/2}\Delta^{1/4}} ,
\end{equation}
consistent with Eq.\ (\ref{hgge}).

If the broadening of the resonance denominator is dominated by electron tunneling, the characteristic range of $\omega$ dominating the integral is determined by $\epsilon_0 - \omega \sim \Gamma_{e}(\omega_-)$, which yields  $\epsilon_0 - \omega \sim (\gamma_e\sqrt{\Delta})^{2/3}$ and a threshold electron tunneling rate of $\Gamma_{e,\mathrm{thres}}=(\gamma_e^2\Delta)^{1/3}\gg \Gamma_{h,\mathrm{thres}}$. Keeping only electron tunneling in the denominator of Eq.\ (\ref{C4}), we obtain the expression 
\begin{equation}
   \left.\frac{\mathrm{d}I}{\mathrm{d}V}\right|_{\mathrm{peak},+} \simeq \frac{4 e^2}{h}\frac{\Gamma_{h,\mathrm{thres}}}{\Gamma_{e,\mathrm{thres}}} \frac{1}{\sqrt{2}}\int_0^{\infty} \frac{  \mathrm{d}x\, x^{5/2}}{(x^3+1/8)^2}
\end{equation}
Performing the integral, we find
\begin{equation}
   \left.\frac{\mathrm{d}I}{\mathrm{d}V}\right|_{\mathrm{peak},+} \simeq \frac{16\pi e^2}{9h}\frac{\Gamma_{h,\mathrm{thres}}}{\Gamma_{e,\mathrm{thres}}} = \frac{8\pi e^2}{9h}\frac{\gamma_h \Delta^{1/6}}{\gamma_e^{2/3}\epsilon_0^{1/2}}  
\end{equation}
consistent with Eq.\ (\ref{eggh}).

\end{document}